\documentclass[%
 reprint,
 amsmath,amssymb,
 aps,
 prl,
 floats,
]{revtex4-1}

\usepackage{subfigure}
\usepackage{graphicx}
\graphicspath{{./}{./images/}{./images_all/}{./images_all/thermal_new/}}
\DeclareGraphicsExtensions{.eps}

\usepackage{dcolumn}
\usepackage{bm}

\begin{document}

\preprint{APS/123-QED}

\title{Binary switching in a symmetric potential landscape}

\author{Kuntal Roy$^{1}$}
\email{royk@vcu.edu}
\author{Supriyo Bandyopadhyay$^1$}
\author{Jayasimha Atulasimha$^2$}
\affiliation{$^1$Dept. of Electrical and Computer Engr., Virginia Commonwealth University, Richmond, VA 23284, USA\\
$^2$Dept. of Mechanical and Nuclear Engr., Virginia Commonwealth University, Richmond, VA 23284, USA}

\date{\today}

\begin{abstract}

The general methodology of binary switching requires tilting of potential landscape along the desired direction of switching. The tilt generates a torque along the direction of switching and the degree of tilt should be sufficient enough to beat thermal agitations with a tolerable error probability. However, we show here that such tilt is not necessary. Considering the full three-dimensional motion, we point out that the \emph{built-in} dynamics can facilitate switching without requiring any asymmetry in potential landscape even in the presence of thermal noise. With experimentally feasible parameters, we theoretically demonstrate such intriguing possibility in electric field-induced magnetization switching of a magnetostrictive nanomagnet.

\end{abstract}

\maketitle

\begin{figure}
\centering
\subfigure[]{\label{fig:switching_asymmetry}\includegraphics[width=2in]
{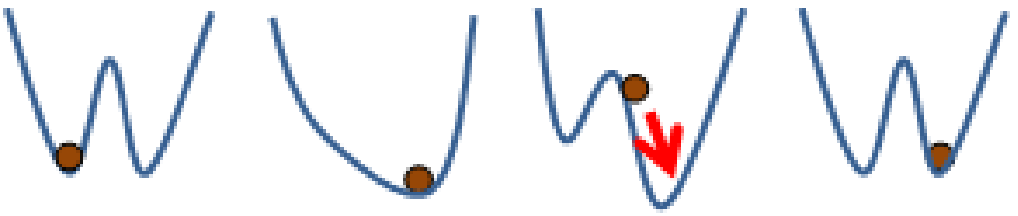}}
\subfigure[]{\label{fig:switching_symmetry}\includegraphics[width=2in]
{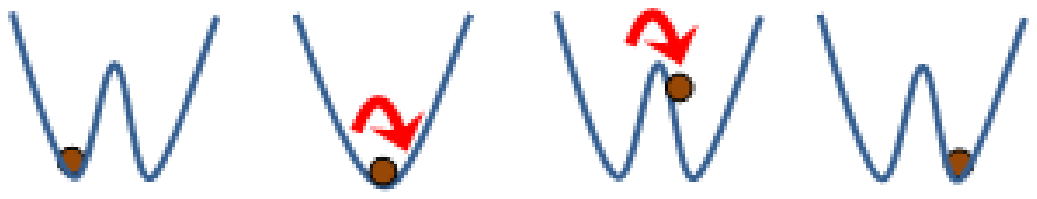}}
\caption{
(a) The potential profile of a bistable switch is made monostable and tilted to switch from one state to another.
(b) The potential profile remains symmetric but switching occurs due to built-in dynamics.}
\end{figure}

Binary switching between two stable states of a bistable element is usually achieved by putting a bias field in the direction of desired state as conceived by Landauer and others~\cite{RefWorks:148, RefWorks:149, RefWorks:144}. The bias field lowers the potential profile along the desired state and thus creates an \emph{asymmetry} in the energy profile. Another external agent inverts the potential profile making it monostable. It is necessary to make the monostable well deep enough to resist thermal fluctuations. Subsequently, the bias field is responsible for switching towards the desired direction when the potential landscape is turned back to bistable. This bias field whether it's a dc field or a field due to dipole coupling with an adjacent element has to overcome the thermal noise such that the switching takes place to the desired direction, however, with a permissible error probability. Such widely accepted methodology is depicted in Fig.~\ref{fig:switching_asymmetry}.

In this Letter, we show that the tilt or asymmetry in potential landscape is not necessary for switching, even in the presence of thermal noise. Our analysis considers the full three-dimensional space during a switching event and depicts the importance of considering the out-of-plane dynamics. We show that the \emph{built-in} dynamics is \emph{sufficient} to ensure error-resiliency. Such consequence cannot be unveiled without the consideration of out-of-plane dynamics. All we need are a sufficiently high magnitude of the external agent inverting the potential landscape and a sufficiently fast ramp rate of the external agent. Basically, the {\it internal dynamics} of the system provides an equivalent asymmetry to cause error-resilient switching. Such intriguing possibility is depicted in Fig.~\ref{fig:switching_symmetry}. The potential energy landscape is never tilted to facilitate switching in the desired direction. However, a torque persists due to out-of-plane motion, which ultimately directs switching in the desired direction.

We theoretically demonstrate such feasibility in a \emph{magnetostrictive} single-domain nanomagnet, where \emph{stress} is the external agent inverting the potential landscape of the nanomagnet. The magnetization of the nanomagnet switches successfully ($>$ 99.99\% probability) \emph{without any requirement of introducing barrier asymmetry} even in the presence of room-temperature thermal fluctuations. The stress can be generated on the magnetostrictive nanomagnet by applying a voltage on a piezoelectric layer elastically coupled to it. Such electric-field induced magnetization switching mechanism has been studied recently because of their potential applications as extreme energy-efficient switches~\cite{roy11}.

\begin{figure}[t]
\centering
\subfigure[]{\label{fig:magnetization_direction}\includegraphics[width=1.6in]
{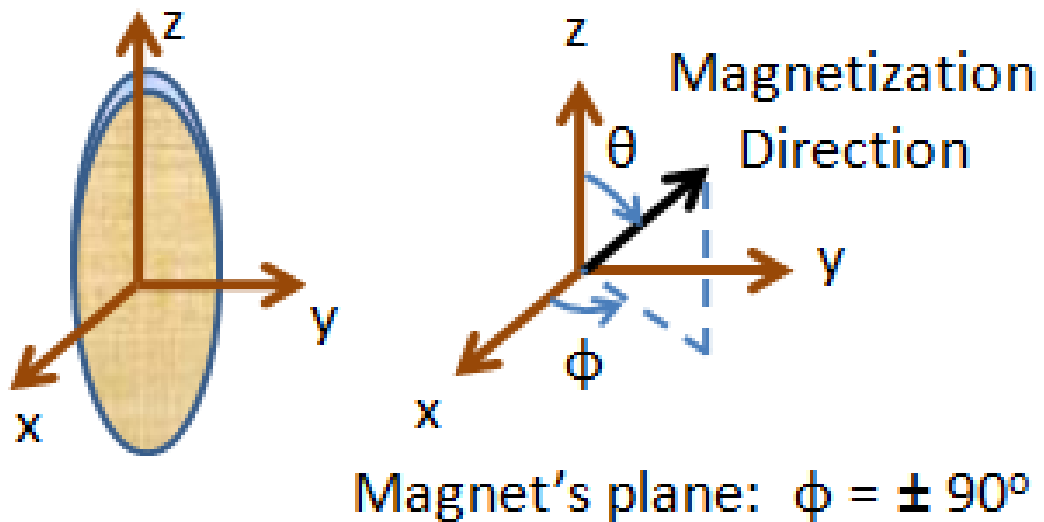}}
\subfigure[]{\label{fig:magnetization_good_quadrants}\includegraphics[width=1.6in]
{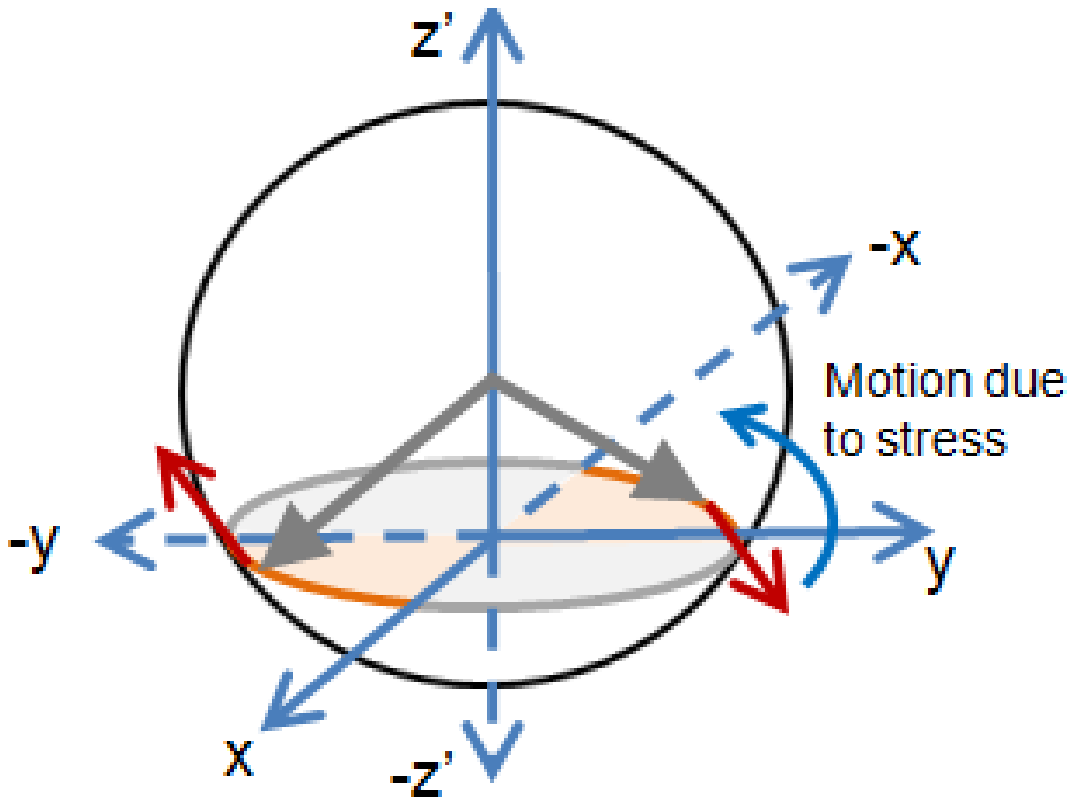}}
\caption{(a) Magnetization in three-dimensional space.
(b) The precessional motion due to applied stress tries to rotate the magnetization out-of-plane in anticlockwise direction while shape anisotropy torque due to small thickness of the nanomagnet tries to bring the magnetization to the magnet's plane ($\phi=\pm90^\circ$). Such counteraction happens only in the quadrants $\phi \in$ ($90^{\circ}$, $180^{\circ}$) and $\phi \in$ ($270^{\circ}$, $360^{\circ}$).}
\end{figure}

Switching dynamics of a single-domain magnetostrictive particle in the presence of thermal fluctuations is modeled using the stochastic Landau-Lifshitz-Gilbert (LLG) equation~\cite{RefWorks:161, RefWorks:186,supplxd} that describes the time-evolution of the magnetization under various torques. There are {\it three} torques to consider here: torque due to shape anisotropy of the nanomagnet, torque due to generated stress anisotropy, and the torque associated with random thermal fluctuations. Consider a magnetostrictive nanomagnet made of polycrystalline Terfenol-D that is shaped like an elliptical cylinder of major axis 100 nm, minor axis 90 nm, and thickness 6 nm. These dimensions ensure that the nanomagnet has a single domain~\cite{RefWorks:133}. Fig.~\ref{fig:magnetization_direction} shows the magnetization in three-dimensional space. We will call the $z$-axis the easy axis, the $y$-axis the in-plane hard axis, and the $x$-axis the out-of-plane hard axis based on the chosen dimensions of the nanomagnet.  The two stable magnetization states are along the $\pm$$z$-axis. In order to switch the magnetization, uniaxial compressive stress is applied in the $z$-direction. To analyze the magnetization dynamics, we will adopt the spherical coordinate system where the magnetization vector is in the radial direction, the polar angle is $\theta$, and the azimuthal angle is $\phi$. Magnetization is on the magnet's plane if $\phi=\pm90^\circ$. 

The total energy of the stressed nanomagnet is the sum of the shape anisotropy and stress anisotropy energy. The torque due to shape and stress anisotropy is derived from the gradient of potential profile and the random thermal torque is added separately in the stochastic LLG dynamics~\cite{supplxd}. We will assume that the magnetization starts from $\theta \simeq 180^\circ$ and the applied stress tries to switch it to $\theta \simeq 0^\circ$. We will use the convention that magnetization's motion is in opposite direction to the torque exerted since the Land\'{e} $g$-factor for electrons is negative.

Dependence of shape-anisotropy energy on azimuthal angle $\phi$ (rather than assuming $\phi=\pm90^\circ$) generates an additional motion of magnetization, which is proportional to $sin(2\phi)\,\hat{\mathbf{e}}_\phi$ and vanishes when $\phi=\pm90^\circ$. As shown in the Fig.~\ref{fig:magnetization_good_quadrants}, the applied stress produces a torque that tries to rotate the magnetization anticlockwise and forces it to reside out-of-plane. As the magnetization is deflected from the plane of the magnet ($\phi=\pm90^\circ$), the additional torque due to shape anisotropy as mentioned above [$\sim$$sin(2\phi)$] would try to bring the magnetization back to its plane. Because of such counteraction, out of the four quadrants for $\phi$, i.e., ($0^{\circ}$, $90^{\circ}$), ($90^{\circ}$, $180^{\circ}$), ($180^{\circ}$, $270^{\circ}$), and ($270^{\circ}$, $360^{\circ}$), the magnetization would be stable in the second quadrant or the fourth quadrant [i.e., ($90^{\circ}$, $180^{\circ}$) or ($270^{\circ}$, $360^{\circ}$)]. Note that $sin(2\phi)$ is \emph{negative} in these two quadrants counteracting the precessional motion due to stress [see Fig.~\ref{fig:magnetization_good_quadrants}]. We would call these two quadrants ``good quadrants'' and the other two (first and third) quadrants ``bad quadrants'', the reasoning behind which would be more prominent onwards. Consideration of the torque due to $\phi$-dependence of shape anisotropy energy is instrumental to the dynamics we present in this Letter. 

\begin{figure}[t]
\centering
\includegraphics[width=2.6in]{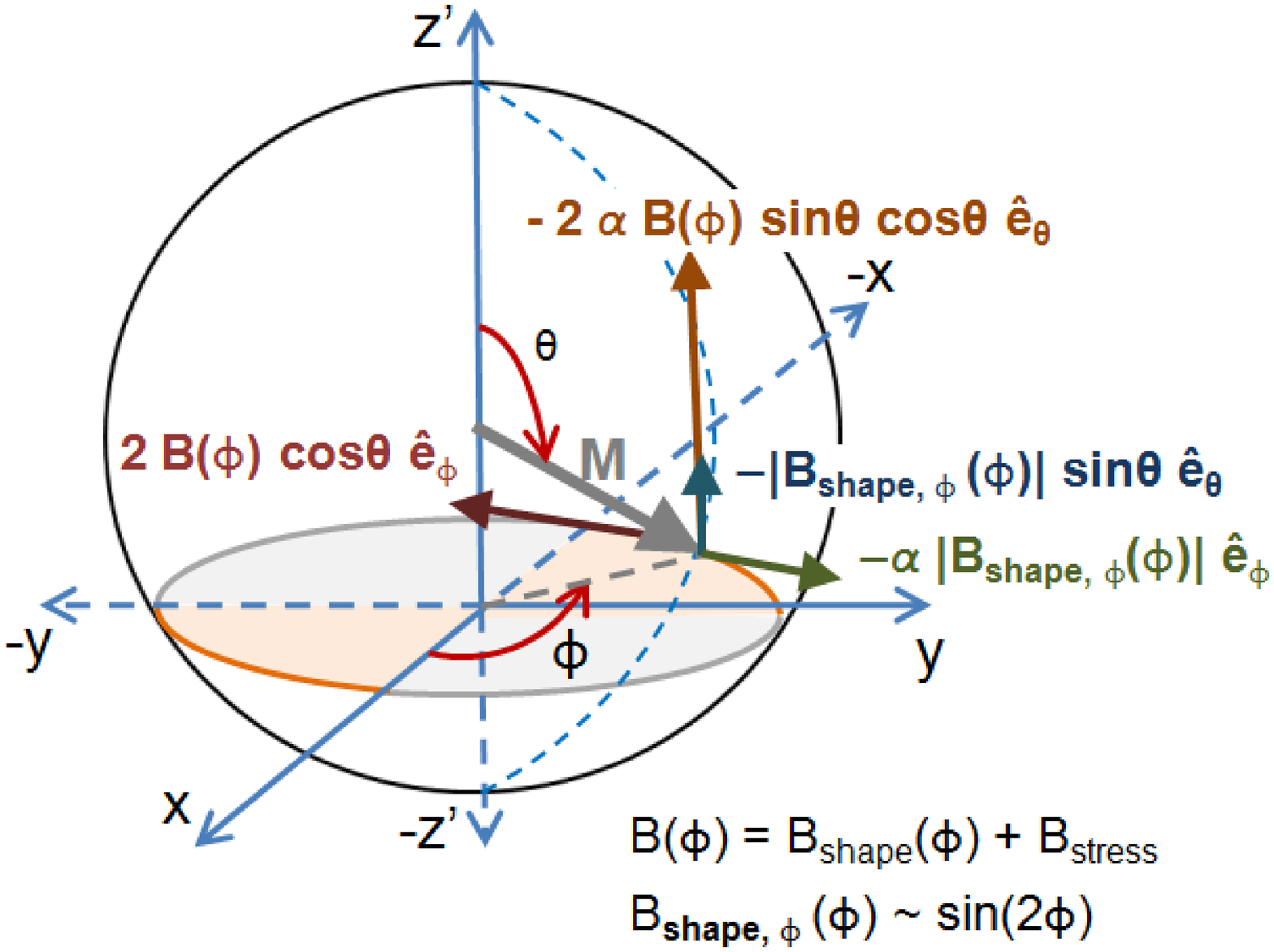}
\caption{\label{fig:motion_illustration} Illustration of magnetization's motion in three-dimensional space under various torques generated due to shape and stress anisotropy alongwith considering the damping of magnetization ($\alpha$ is the phenomenological damping constant). Note that the dependence of shape anisotropy energy on $\phi$ has generated two additional motions $-|B_{shape,\phi}(\phi)|sin\,\theta\,\hat{\mathbf{e}}_\theta$ and $-\alpha |B_{shape,\phi}(\phi)|\,\hat{\mathbf{e}}_\phi$. The quadrant $\phi \in$ ($90^{\circ}$, $180^{\circ}$) is chosen for illustration; choice of the other good quadrant $\phi \in$ ($270^{\circ}$, $360^{\circ}$) is analogous.}
\end{figure}

We will now describe the motion of magnetization intuitively under various torques originating from shape and stress anisotropy as shown in the Fig.~\ref{fig:motion_illustration}. We intend the motion of magnetization to be along the -$\hat{\mathbf{e}}_\theta$ direction since we are switching magnetization from $\theta \simeq 180^\circ$ towards $\theta \simeq 0^\circ$. The precessional motion of magnetization due to torque generated by the applied stress is in the +$\hat{\mathbf{e}}_\phi$ direction, but the damping of magnetization generates an additional motion, which is perpendicular to both the direction of magnetization ($\hat{\mathbf{e}}_r$) and +$\hat{\mathbf{e}}_\phi$, i.e., in -$\hat{\mathbf{e}}_\theta$ direction. These two motions are depicted as $2B(\phi)cos\theta\,\hat{\mathbf{e}}_\phi$ and $-2\alpha B(\phi)sin\theta cos\theta\,\hat{\mathbf{e}}_\theta$, respectively in Fig.~\ref{fig:motion_illustration}, where $\alpha$ is the damping constant~\cite{RefWorks:161}, $B(\phi)=B_{shape}(\phi)+B_{stress}$, $B_{shape}(\phi)$ is the $\phi$-dependent strength of shape anisotropy energy, and $B_{stress}$ is the strength of the stress anisotropy energy~\cite{supplxd}. The quantity $B_{stress}$ is negative and it must beat the shape anisotropy energy for switching to get started. Mathematically, note that both the quantities $B(\phi)$ and $cos\,\theta$ are negative in the interval $180^\circ \geq \theta \geq 90^\circ$. Hence, magnetization switches towards its desired direction due to the applied stress. However, this damped motion in -$\hat{\mathbf{e}}_\theta$ direction is weak because of the multiplicative factor, $\alpha$, which is usually much less than one (e.g., $\alpha$=0.1 for Terfenol-D). 

As magnetization rotates out-of-plane due to applied stress, and stays in the ``good quadrants'' for $\phi$ [i.e., ($90^{\circ}$, $180^{\circ}$) or ($270^{\circ}$, $360^{\circ}$)] as described earlier [see Fig.~\ref{fig:magnetization_good_quadrants}], it generates a motion of magnetization in the -$\hat{\mathbf{e}}_\theta$ ($-\hat{\mathbf{e}}_\phi \times \hat{\mathbf{e}}_r$) direction due to $\phi$-dependence of shape anisotropy energy. Subsequently, a damped motion is generated too in the -$\hat{\mathbf{e}}_\phi$ ($\hat{\mathbf{e}}_r \times -\hat{\mathbf{e}}_\theta$) direction. These two motions are depicted as $-|B_{shape,\phi}(\phi)|sin\theta\,\hat{\mathbf{e}}_\theta$ and $-\alpha |B_{shape,\phi}(\phi)|\,\hat{\mathbf{e}}_\phi$, respectively in Fig.~\ref{fig:motion_illustration}, where $B_{shape,\phi}(\phi) \sim sin(2\phi)$. Note that in the good quadrants for $\phi$, $B_{shape,\phi}(\phi)$ is negative. Thus, keeping the magnetization out-of-plane of the magnet in \emph{good quadrants} is beneficial in switching the magnetization in its desired direction. In case the magnetization resides out-of-plane but in the \emph{bad quadrants}, it would have resisted the motion of magnetization in its desired direction of switching. A higher magnitude of stress keeps the magnetization more out-of-plane in \emph{good quadrants} due to precessional motion, however, the damped motion $-\alpha |B_{shape,\phi}(\phi)|\,\hat{\mathbf{e}}_\phi$ tries to bring magnetization back towards the magnet's plane. As these two motions counteract each other, magnetization keeps moving in the -$\hat{\mathbf{e}}_\theta$ direction and eventually reaches at the $x$-$y$ plane defined by $\theta=90^\circ$.

\begin{figure}
\centering
\subfigure[]{\label{fig:stress_cycle}\includegraphics[width=3.4in]
{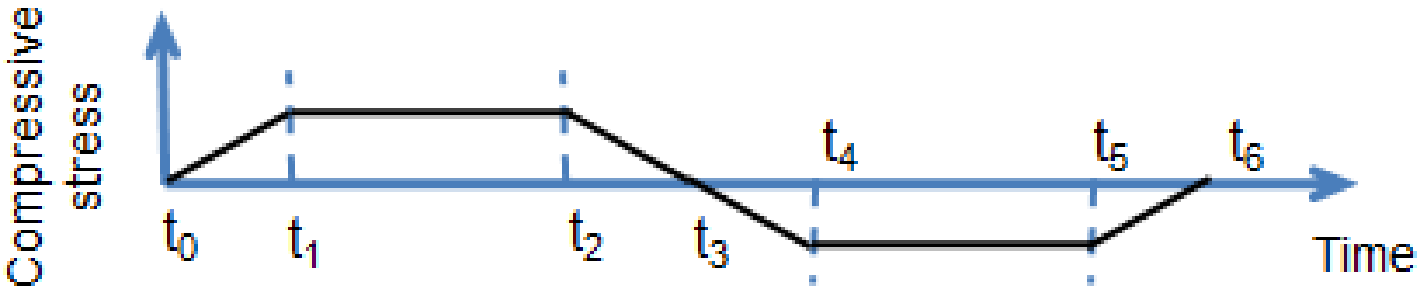}}
\subfigure[]{\label{fig:stress_cycle_magnetization}\includegraphics[width=3.4in]
{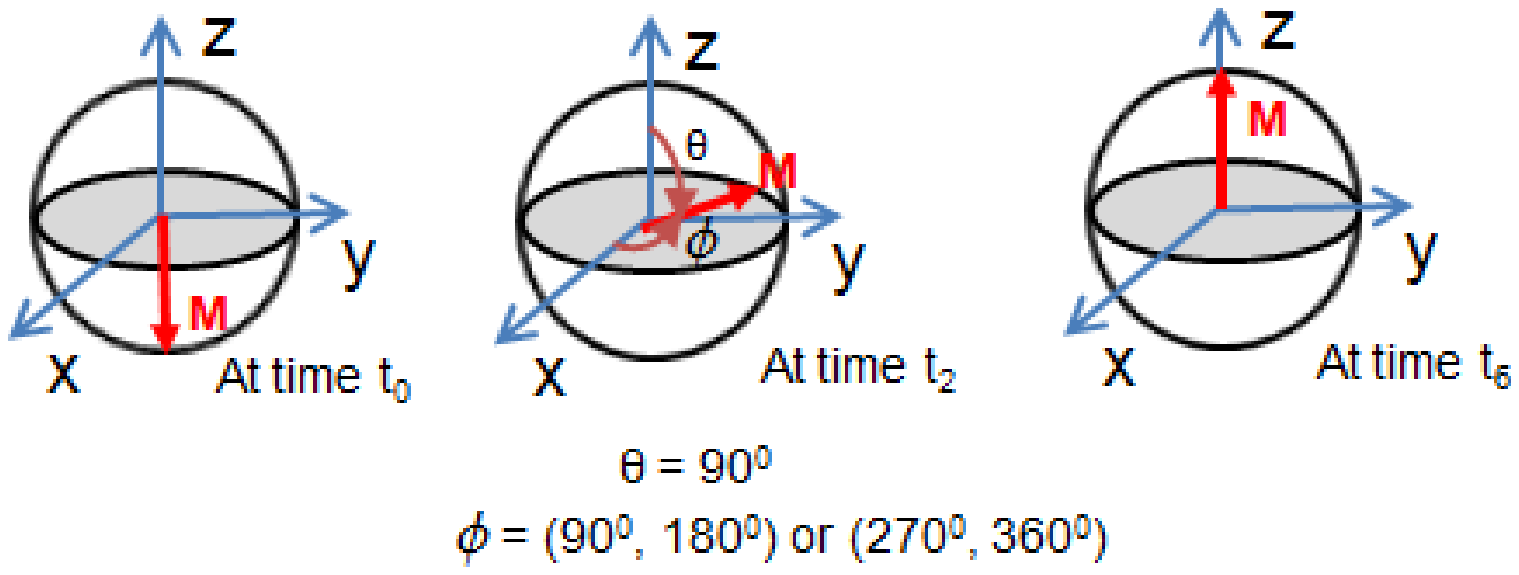}}
\subfigure[]{\label{fig:stress_cycle_energy_profile}\includegraphics[width=3.4in]
{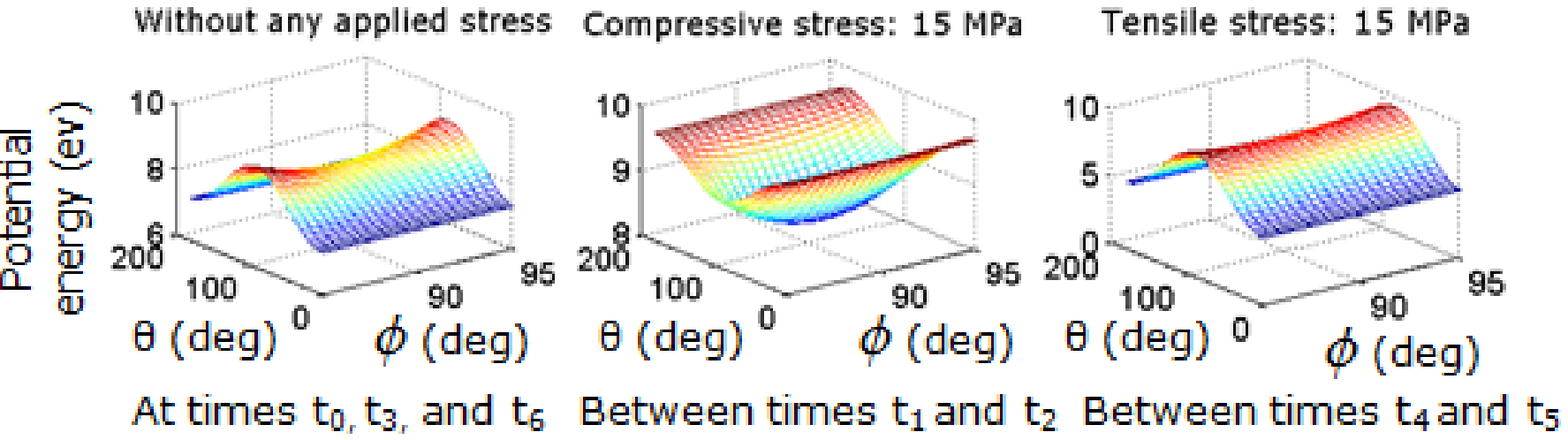}}
\caption{\label{fig:stress_cycle_all} (a) Stress-cycle on the magnetostrictive nanomagnet, 
(b) Magnetization directions at different instants of time, and (c) Potential landscapes of the magnetostrictive nanomagnet in relaxed, compressively stressed, and expansively stressed conditions.}
\end{figure}

The stress-cycle alongwith the energy profiles and magnetization directions at different instants of time is shown in the Fig.~\ref{fig:stress_cycle_all}. At time $t_0$, the magnetization direction is along the easy axis $\theta \simeq 180^\circ$ and the potential landscape of the nanomagnet is unperturbed by stress. The potential profile of the magnet is symmetric in both $\theta$- and $\phi$-space with two degenerate minima at $\theta = 0^{\circ}$, 180$^{\circ}$ and a maximum at $\theta = 90^{\circ}$ in $\theta$-space. The anisotropy in the barrier is due to shape anisotropy only, which is $\sim$44 kT at room-temperature using the magnet's dimensions and material parameters of Terfenol-D~\cite{supplxd}. Note that the barrier height separating the two stable states ($\theta$ = $0^\circ$ and $180^\circ$) is meant when the magnetization resides in-plane (i.e., $\phi=\pm90^\circ$) of the magnet. The barrier goes higher when the magnetization is deflected from $\phi=90^\circ$ as shown in the Fig.~\ref{fig:stress_cycle_energy_profile} [at time $t_0$]. The barrier is highest when the magnetization points along the out-of-plane direction ($\phi$ = $0^\circ$ or $180^\circ$). Magnetization can start from any angle $\phi_{initial} \in (0^\circ, 360^\circ)$ in the presence of thermal noise~\cite{supplxd}.

As a compressive stress is ramped up on the nanomagnet between time instants $t_0$ and $t_1$, the potential landscape in $\theta$-space becomes monostable near $\phi=\pm90^\circ$ provided a sufficient stress is applied. The potential barrier near $\phi=0^\circ$ or $180^\circ$ may not become monostable in $\theta$-space since barrier height is high therein, however, that is not necessary for switching. Since application of stress rotates the magnetization in $\phi$-direction, it can eventually come near $\phi=\pm90^\circ$ and starts switching from $\theta \simeq 180^\circ$ towards $\theta=90^\circ$. The minimum energy position between time instants $t_1$ and $t_2$ is at ($\theta=90^\circ$, $\phi=\pm90^\circ$). From Fig.~\ref{fig:stress_cycle_energy_profile}, we can see that the potential profile at time instant $t_1$ is still \emph{symmetric}.

\begin{figure}
\centering
\subfigure[]{\label{fig:good_quadrant_0_90}\includegraphics[width=1.6in]
{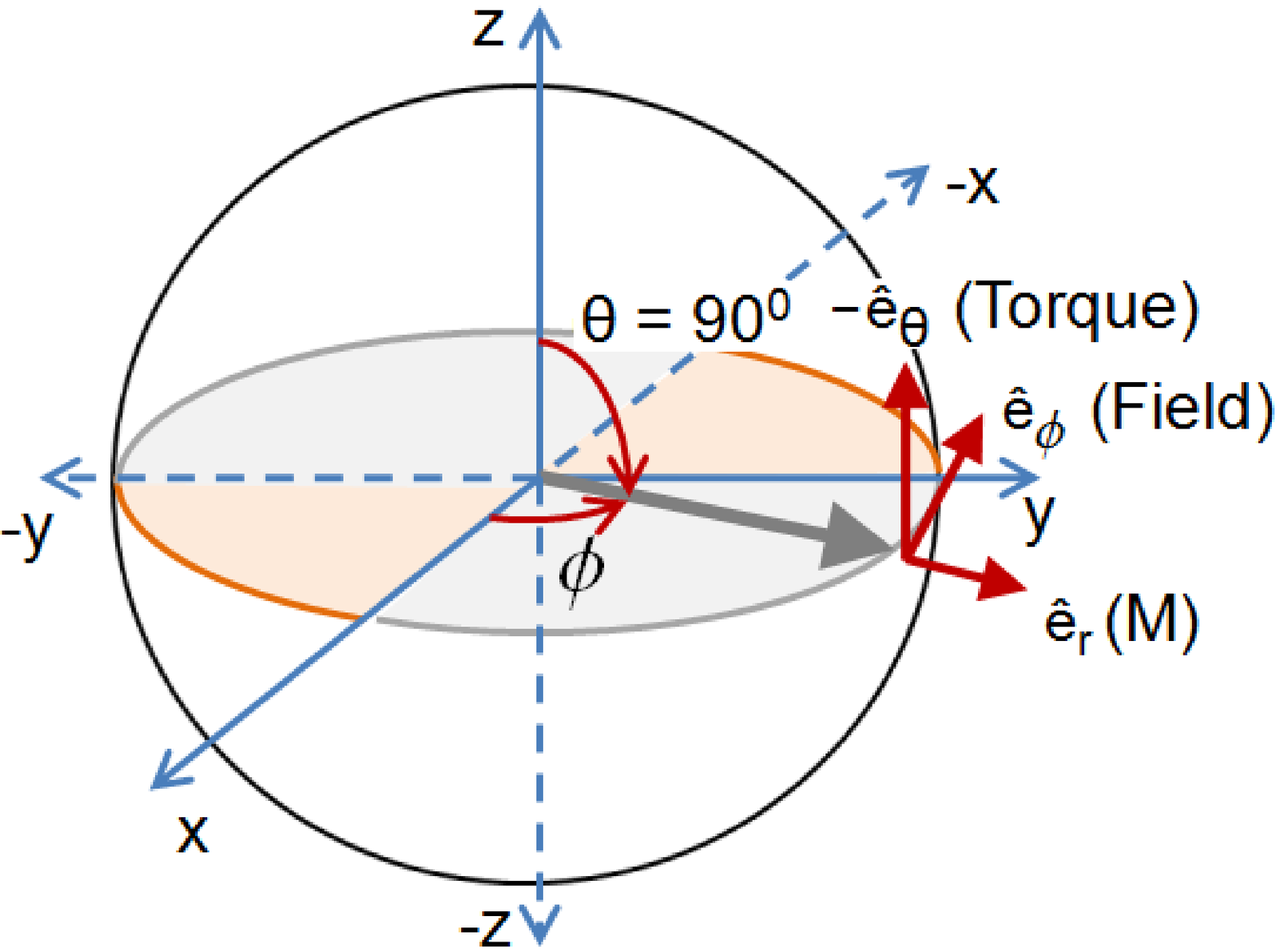}}
\subfigure[]{\label{fig:bad_quadrant_90_180}\includegraphics[width=1.6in]
{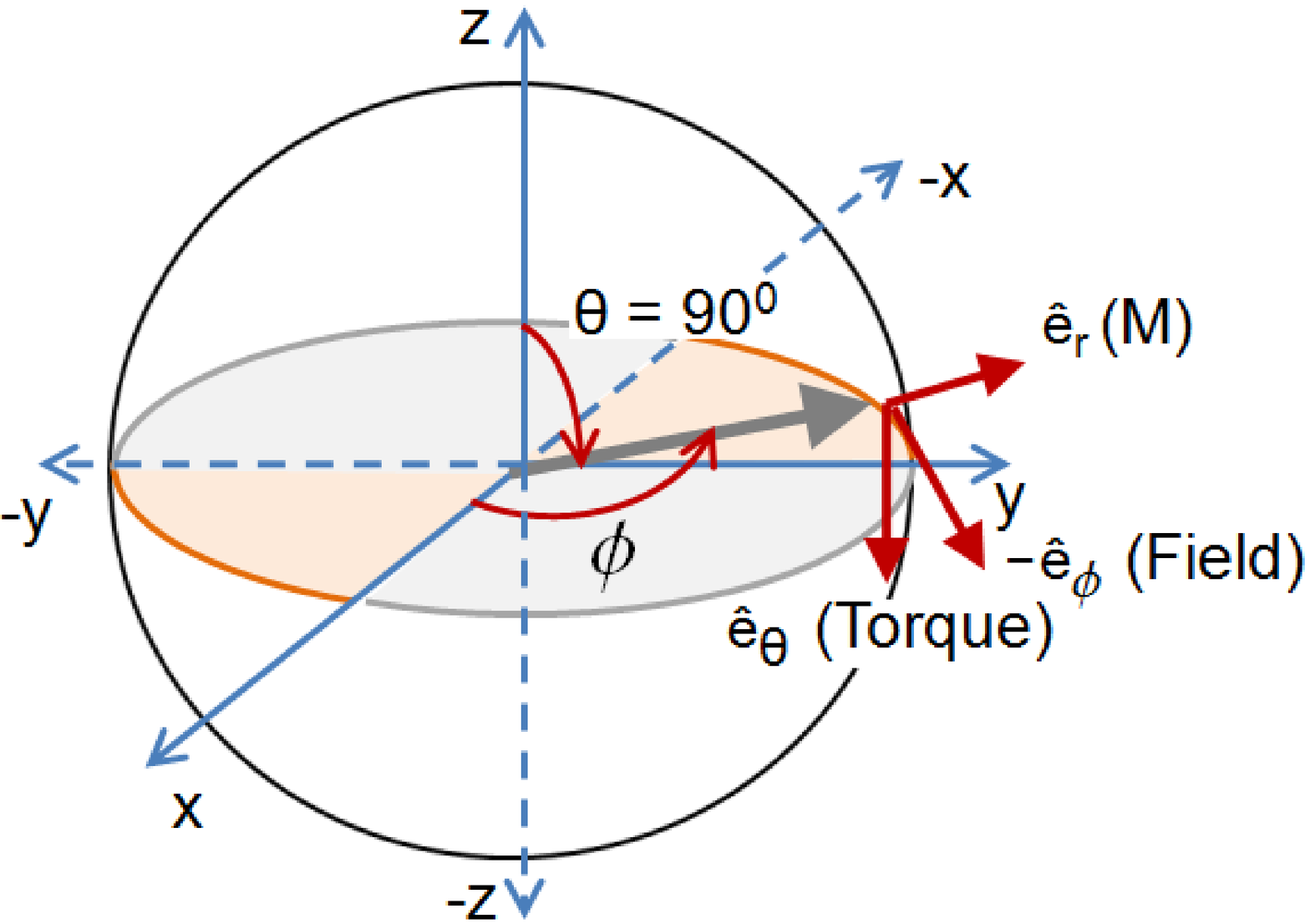}}
\caption{\label{fig:good_bad_quadrants} Field and torque on the magnetization, \textbf{M} when it comes on the $x$-$y$ plane ($\theta \simeq 90^\circ$). The field always tries to keep the magnetization in-plane ($\phi=\pm90^\circ$) of the nanomagnet. Magnetization started from $\theta \simeq 180^\circ$. (a) $\phi \in (0^\circ, 90^\circ)$, (b) $\phi \in (90^\circ, 180^\circ)$. Magnetization can traverse towards its destination $\theta \simeq 0^\circ$ for the case (b), while it backtracks towards $\theta \simeq 180^\circ$ for the case (a). Note that the motion of magnetization is opposite to the direction of torque exerted on it. The quadrant $\phi \in (180^\circ, 270^\circ)$ is analogous to the case (a), while the quadrant $\phi \in (270^\circ, 360^\circ)$ is analogous to the case (b). }
\end{figure}

Upon reaching at $\theta=90^\circ$, if magnetization stays in the ``good quadrants'' for $\phi$ [i.e., $(90^\circ,180^\circ)$ or $(270^\circ,360^\circ)$], then the torque on the magnetization is in the correct direction so that it can traverse towards $\theta \simeq 0^\circ$. (See Fig.~\ref{fig:good_bad_quadrants}.) This once again signifies the merit of terminology (good or bad) used for the four quadrants of $\phi$. Note that at $\theta=90^\circ$ (i.e., $cos\,\theta=0$), the effect of stress on the magnetization rotation has diminished completely. The only two motions that are active at that point are $-|B_{shape,\phi}(\phi)|sin\,\theta\,\hat{\mathbf{e}}_\theta$ and $-\alpha |B_{shape,\phi}(\phi)|\,\hat{\mathbf{e}}_\phi$ (see Fig.~\ref{fig:motion_illustration}). Since $\alpha \ll 1$, magnetization quickly gets out of the $\theta = 90^\circ$ position and as the magnetization vector gets deflected from $\theta=90^\circ$ towards $\theta=0^\circ$, the effect of stress comes into play.

As the magnetization leaves from $\theta=90^\circ$ towards $\theta \simeq 0^\circ$ and stress is ramped down at time $t_2$ (see Fig.~\ref{fig:stress_cycle_all}), the torque due to stress tries to rotate the azimuthal angle $\phi$ of magnetization \emph{clockwise} rather than anticlockwise (mathematically note that $cos\,\theta$ is positive for $90^\circ \geq \theta \geq 0^\circ$ and $B(\phi)$ is still negative when stress has not been brought down significantly, i.e., still $|B_{stress}| > |B_{shape}(\phi)|$). For a \emph{slow ramp-rate} this rotation may be considerable and magnetization can stray into bad quadrants [$(0^\circ,90^\circ)$ or $(180^\circ,270^\circ)$]. Moreover, thermal fluctuations can aggravate the scenario. Switching may impede and magnetization vector can backtrack towards where it started ($\theta \simeq 180^\circ$) causing a switching failure. In this way, switching failure may happen even after the magnetization has crossed the hard axis ($\theta = 90^\circ$) towards its destination ($\theta \simeq 0^\circ$). This is why it does need a fast enough ramp rate during the ramp-down phase of stress. Note that when the potential landscape gets inverted from its unperturbed position (between the instants of time $t_0$ and $t_1$ in Fig.~\ref{fig:stress_cycle_all}), the effect of thermal fluctuations does not matter since it can only delay the magnetization to come at $\theta=90^\circ$. 

\begin{figure}
\centering
\includegraphics[width=2.2in]{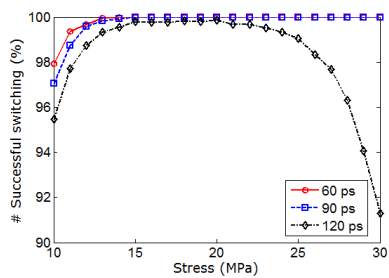}
\caption{\label{fig:thermal_stress_success_ramp_time} Percentage of
successful switching events at room temperature (300 K) in a Terfenol-D/PZT multiferroic when subjected to stress between 10 MPa and 30 MPa. The critical stress at which switching becomes $\sim$100\% successful increases with ramp duration. However, at high ramp duration (e.g., 120 ps), we may not achieve $\sim$100\% switching probability for any values of stress.}
\end{figure}

Straying into a bad quadrant for azimuthal angle $\phi$ [$(0^\circ,90^\circ)$ or $(180^\circ,270^\circ)$] during the ramp-down phase does not necessarily mean that the magnetization would fail to switch. Particularly if magnetization is close to its destination $\theta \simeq 0^\circ$, the shape anisotropy torque alongwith the \emph{negative} stress anisotropy torque (since stress is \emph{reversed}) would have enough control to bring the magnetization back to the other good quadrant for $\phi$ and complete the switching. Thus, there may be ripples appearing in the magnetization dynamics at the end of switching, which is because of the transition of azimuthal angle $\phi$ between two good quadrants through one bad quadrant~\cite{supplxd}. 

Therefore, we require the following two criteria for successful switching without creating any asymmetry in the potential landscape even in the presence of thermal fluctuations: (1) a high enough stress that keeps the magnetization more out-of-plane inside the good quadrants; and (2) a fast enough ramp rate that reduces the possibility of backtracking of magnetization while it is crossing (or even after crossing) the hard axis ($\theta = 90^\circ$) towards its destination. A high stress and a fast ramp rate also increase the switching speed and surpass the detrimental effects of thermal fluctuations. 

Fig.~\ref{fig:thermal_stress_success_ramp_time} shows the switching probability for 
different values of stress (10-30 MPa) and ramp durations (60 ps, 90 ps, 120 ps) at room temperature (300 K). A moderately large number (10,000) of simulations were performed for each value of stress and ramp duration to generate these results. Initial angle distributions at 300 K for both $\theta$ and $\phi$ are taken into account during simulations. The minimum stress needed to switch the magnetization at 0 K is $\sim$5 MPa, but the minimum stress needed to ensure switching at 300 K is $\sim$14 MPa for 60 ps ramp duration and $\sim$17 MPa for 90 ps ramp duration. For  120 ps ramp duration, $\sim$100\% success probability is unattainable for any values of stress since thermal fluctuations have higher latitude to divert the magnetization in wrong direction while stress is ramped down; at higher stress values accompanied by a high ramp duration, there occurs higher out-of-plane excursion pushing the magnetization in bad quadrants, which further aggravates the error probability.

In conclusion, we have shown that binary switching in a symmetric potential landscape is feasible, even in the presence of thermal noise. We have theoretically demonstrated such possibility in successful magnetization reversal of a single magnetostrictive particle. The out-of-plane dynamics plays the crucial role to ensure such error-resilient switching. All we need are a sufficiently fast ramp rate and a sufficiently high magnitude of the stress applied on the magnetostrictive particle, which are experimentally feasible. 

This work was supported by the US National Science Foundation under the NEB 2020 grant ECCS-1124714 and by the Semiconductor Research Corporation under NRI task 2203.001.


%

\end{document}


\maketitle

In this supplementary section, we provide further details of the analysis and some additional results.

\section{Magnetization dynamics of a magnetostrictive nanomagnet in the 
presence of thermal noise: Solution of the stochastic Landau-Lifshitz-Gilbert 
equation}

\begin{figure}
\centering
\includegraphics[width=3in]{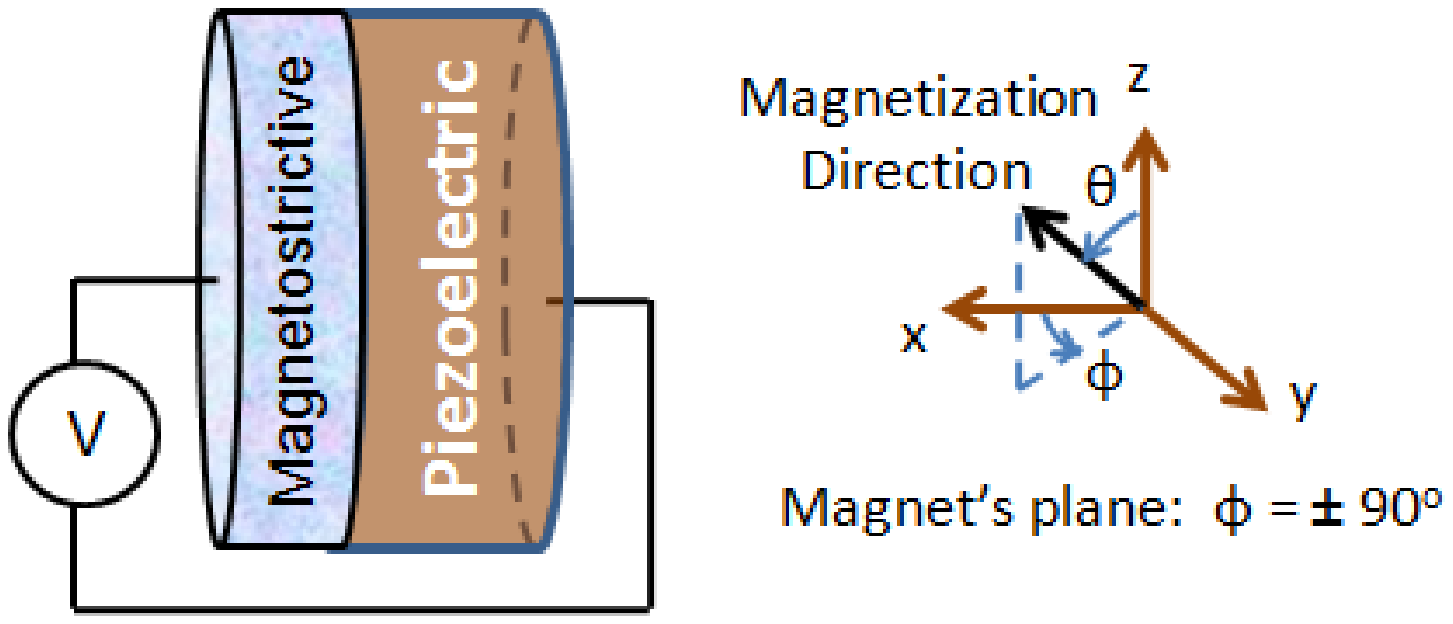}
\caption{\label{fig:multiferroic} An elliptical two-phase multiferroic nanomagnet stressed with an applied 
voltage. The voltage generates strain in the piezoelectric layer which is mostly transferred to 
the magnetostrictive layer. The resulting stress 
in the magnetostrictive layer favors aligning the magnetization vector in the plane defined by the in-plane hard 
axis ($y$-direction) and the out-of-plane hard axis ($x$-direction) rather than along the easy 
axis ($z$-direction)~\cite{roy11}.}
\end{figure}

Consider an isolated nanomagnet in the shape of an elliptical cylinder whose elliptical cross section lies in the $y$-$z$ plane with its 
major axis aligned along the $z$-direction and minor axis along the $y$-direction. (See Fig.~\ref{fig:multiferroic}.) The dimension of the major axis is $a$, that of the minor axis is $b$, and the thickness is $l$. The volume of the nanomagnet is $\Omega=(\pi/4)a b l$. Let $\theta(t)$ be the angle subtended by the magnetization axis with the +$z$-axis at any instant of time $t$ and $\phi(t)$ be the angle between the +$x$-axis and the projection of the magnetization axis on the $x$-$y$ plane. Thus, $\theta(t)$ is the polar angle and $\phi(t)$ is the azimuthal angle. 
Note that when $\phi =\pm 90^{\circ}$, the magnetization vector lies in the plane of the magnet. Any deviation from $\phi = \pm 90^{\circ}$ 
corresponds to out-of-plane excursion.

The total energy of the single-domain, magnetostrictive, polycrystalline
nanomagnet, subjected to uniaxial stress along the easy axis (major axis of the ellipse) is the sum of the uniaxial shape anisotropy energy and the uniaxial stress anisotropy energy~\cite{RefWorks:157}:
\begin{equation}
E(t) = E_{SHA}(t) + E_{STA}(t),
\end{equation}
where $E_{SHA}(t)$ is the uniaxial shape anisotropy energy and $E_{STA}(t)$ is the uniaxial 
stress anisotropy energy at time $t$. The former is given by~\cite{RefWorks:157}
\begin{equation}
E_{SHA}(t) = (\mu_0/2) M_s^2 \Omega N_d(t)
\label{shape-anisotropy}
\end{equation}
where $M_s$ is the saturation magnetization and $N_d(t)$ is the demagnetization factor expressed 
as~\cite{RefWorks:157} 
\begin{equation}
N_d(t) = N_{d-zz} cos^2\theta(t) + N_{d-yy} sin^2\theta(t) \ sin^2\phi(t)  + N_{d-xx} sin^2\theta(t) \, cos^2\phi(t)
\end{equation}
with $N_{d-zz}$, $N_{d-yy}$, and $N_{d-xx}$ being the components of the
demagnetization factor along the $z$-axis, $y$-axis, and $x$-axis, respectively. The parameters $N_{d-zz}$, 
$N_{d-yy}$, and $N_{d-xx}$ depend on the shape and dimensions of the 
nanomagnet and are determined from the prescription in Ref.~\cite{RefWorks:402}.

Note that uniaxial shape anisotropy will favor lining up the magnetization along the major axis 
($z$-axis) by minimizing $E_{SHA}(t)$, which is why we will call the major axis the ``easy axis'' 
and the minor axis ($y$-axis) the ``in-plane hard axis'' of the magnet. The $x$-axis will therefore be the ``out-of-plane hard axis'' 
of the magnet and it is ``harder'' than the in-plane one since the thickness is much smaller than the magnet's lateral
 dimensions (i.e., $l << a, b$). By constraining the nanomagnet from expanding or contracting in the $y$-direction, we can generate 
uniaxial stress along the $z$-axis (easy axis). An appropriate way to do this is to deposit the magnetostrictive layer on a piezoelectric layer to make a 2-phase multiferroic. By applying a potential across the piezoelectric layer, we generate a strain in it which is transferred to 
the magnetostrictive layer by elastic coupling. This generates uniaxial stress in the latter layer.
It is the preferred technique for generating stress since it is electrical in nature and dissipates very little amount of energy \cite{roy11}.

The stress anisotropy energy is given by~\cite{RefWorks:157}
\begin{equation}
E_{STA}(t) = - (3/2) \lambda_s \sigma(t) \Omega \, cos^2\theta(t)
\label{eq:e_sta}
\end{equation}
where $(3/2) \lambda_s$ is the magnetostriction coefficient of the nanomagnet and $\sigma(t)$ is the stress 
at an instant of time $t$. 
Note that a positive $\lambda_s \sigma(t)$ product will favor alignment of the magnetization along 
the major axis ($z$-axis), while a negative $\lambda_s \sigma(t)$ product will favor alignment 
along the minor axis ($y$-axis), because that will minimize $E_{STA}(t)$. In our convention, 
a compressive stress is negative and tensile stress is positive. Therefore, in a material 
like Terfenol-D that has positive $\lambda_s$, a compressive stress will favor alignment 
along the minor axis, and tensile along the major axis. The situation will be opposite 
with nickel and cobalt that have negative $\lambda_s$.
 
At any instant of time $t$, the total energy of the nanomagnet can be expressed as 
\begin{equation}
E(t) = E(\theta(t),\phi(t),\sigma(t)) = B(t) sin^2\theta(t) + C(t)
\end{equation}
where 
\begin{subequations}
\begin{align}
B(t) &= B_{shape}(t) + B_{stress}(t) \\
B_{shape}(t) &= B_{shape}(\phi(t)) = \cfrac{\mu_0}{2} \, M_s^2 \Omega \left\lbrack N_{d-xx} cos^2\phi(t) + N_{d-yy} sin^2\phi(t) - N_{d-zz}\right\rbrack \\
B_{stress}(t) &= (3/2) \lambda_s \sigma(t) \Omega \\
C(t) &= \cfrac{\mu_0}{2} M_s^2 \Omega N_{d-zz} - (3/2) \lambda_s \sigma(t) \Omega.
\end{align}
\end{subequations}
Note that $B_{stress}(t)$ has the same sign as the $\lambda _s \sigma(t)$ product.
It will be negative if we use stress to rotate the magnetization from the easy axis ($z$-direction) to the plane defined by the in-plane hard 
axis ($y$-direction) and the out-of-plane hard axis ($x$-direction).

In the macrospin approximation, the magnetization \textbf{M}(t) of the nanomagnet has a constant magnitude at any given temperature but a variable direction,
so that we can represent it by the vector of unit norm $\mathbf{n_m}(t) =\mathbf{M}(t)/|\mathbf{M}| = \mathbf{\hat{e}_r}$ where 
$\mathbf{\hat{e}_r}$ is the unit vector in the radial direction in spherical coordinate system represented by ($r$,$\theta$,$\phi$). 
This means that the magnetization vector is in the radial direction, and the polar and azimuthal angles $\left ( \theta(t), \phi(t) \right)$
will specify its orientation at any given time $t$.
The  unit vectors in the $\theta$- and $\phi$-directions are denoted by $\mathbf{\hat{e}_\theta}$ and $\mathbf{\hat{e}_\phi}$,
 respectively. 

The gradient of potential energy at any particular instant of time $t$ is given by
\begin{equation}
\nabla E(t) = \nabla E(\theta(t),\phi(t),\sigma(t)) = \cfrac{\partial E(t)}{\partial \theta(t)} \, \mathbf{\hat{e}_\theta} + \cfrac{1}{sin\theta(t)} \,\cfrac{\partial E(t)}{\partial \phi(t)} \, \mathbf{\hat{e}_\phi} 
\end{equation}
\noindent
where
\begin{equation}
\cfrac{\partial E(t)}{\partial \theta(t)} = 2 B(t) sin\theta (t) cos\theta(t)
\end{equation}
\noindent
and
\begin{eqnarray}
\cfrac{\partial E(t)}{\partial \phi(t)} &=& -\frac{\mu_0}{2} \, M_s^2 \Omega (N_{d-xx}-N_{d-yy}) 
sin(2\phi(t)) sin^2\theta (t) \nonumber \\ 
&=& - B_{shape,\phi}(t) \, sin^2\theta (t).
\end{eqnarray}
\noindent
Here 
\begin{eqnarray}
B_{shape,\phi}(t)&=&B_{shape,\phi}(\phi(t))=\cfrac{\mu_0}{2} \, M_s^2 \Omega (N_{d-xx}-N_{d-yy}) sin(2\phi(t)). \label{eq:B0e}
\label{definitions}
\end{eqnarray}

The torque acting on the magnetization per unit volume due to shape and stress anisotropy is
\begin{eqnarray}
\mathbf{T_E} (t) &=& - \mathbf{n_m}(t) \times \nabla E(\theta(t),\phi(t),\sigma(t)) \nonumber\\
&=& - \mathbf{\hat{e}_r} \times \left\lbrack 2 B(t) sin\theta(t) cos\theta(t) \,\mathbf{\hat{e}_\theta} - B_{shape,\phi}(t)  sin\theta (t) \,\mathbf{\hat{e}_\phi} \right\rbrack  
 \nonumber\\
&=& - 2 B(t) sin\theta(t) cos\theta(t) \,\mathbf{\hat{e}_\phi}  - B_{shape,\phi}(t) sin\theta (t) \,\mathbf{\hat{e}_\theta}, 							 
\label{eq:stress_torque}
\end{eqnarray}
\noindent
whereas the torque due to thermal fluctuations is treated via a random magnetic field $\mathbf{h}(t)$ expressed as 
\begin{equation}
\mathbf{h}(t) = h_x(t)\mathbf{\hat{e}_x} + h_y(t)\mathbf{\hat{e}_y} + h_z(t)\mathbf{\hat{e}_z}
\end{equation}
\noindent
where $h_x(t)$, $h_y(t)$, and $h_z(t)$ are the three components of $\mathbf{h}(t)$ in $x$-, $y$-, and $z$-direction, respectively 
in Cartesian coordinates. We will assume the following properties of the random field, $\mathbf{h}(t)$~\cite{RefWorks:186}.
\begin{itemize}
	\item The process $\mathbf{h}(t)$ is \emph{stationary}.
	\item The distribution of the quantities $h_x(t)$, $h_y(t)$, and $h_z(t)$ is normal (Gaussian) with means equal to zero, i.e. $\langle h_i(t) \rangle = 0$ where $i=x,y,z$.
	\item The quantities $h_i(t)$ and $h_j(t')$ (where $t'-t=\pm\Delta$ and $i\neq j$) are correlated only for time intervals 
$\Delta$, which is much shorter than the time it takes for the magnetization vector to rotate appreciably. Furthermore
	\begin{equation}
		\langle h_i(t) h_j(t) \rangle = U \delta_{ij} \delta(\Delta) \qquad (i,j=x,y,z)
	\end{equation}
	\noindent
	where $U=\frac{2 \alpha KT}{|\gamma| (1+\alpha^2) M_V}$~\cite{RefWorks:186, RefWorks:422}, $\alpha$ is the dimensionless phenomenological Gilbert damping constant, $\gamma = 2\mu_B \mu_0/\hbar$ is the gyromagnetic ratio for electrons and is equal to $2.21\times 10^5$ (rad.m).(A.s)$^{-1}$, $\mu_B$ is the Bohr magneton, and $M_V=\mu_0 M_s\Omega$.
	\item The statistical properties of the quantities $h_x(t)$, $h_y(t)$, and $h_z(t)$ are isotropic.
\end{itemize}
\noindent
Accordingly, the random thermal field can be expressed as
\begin{equation}
h_i(t) = \sqrt{\frac{2 \alpha kT}{|\gamma| (1+\alpha^2) M_V \Delta t}} \; G_{(0,1)}(t) \qquad (i=x,y,z)
\label{eq:ht}
\end{equation}
\noindent
where $1/\Delta t$ is proportional to the attempt frequency of the random thermal field. Consequently,
$\Delta t$ should be the simulation time-step used to solve the coupled LLG equations numerically and $G_{(0,1)}(t)$ is a Gaussian distribution with zero mean and unit standard deviation~\cite{RefWorks:388}. The simulation time-step $\Delta t$ should be selected small enough so that decreasing that step further does not make any significant difference in the results. Accordingly, the thermal torque can be written as
\begin{equation}
\mathbf{T_{TH}}(t) = M_V\,\mathbf{n_m}(t) \times \mathbf{h}(t) = P_\theta(t)\,\mathbf{\hat{e}_\phi} - P_\phi(t)\,\mathbf{\hat{e}_\theta}
\end{equation}
\noindent
where
\begin{eqnarray}
P_\theta(t) &=& M_V\left\lbrack h_x(t)\,cos\theta(t)\,cos\phi(t) + h_y(t)\,cos\theta(t)sin\phi(t) - h_z(t)\,sin\theta(t) \right\rbrack \label{eq:thermal_parts_theta}\\
P_\phi(t) &=& M_V\left\lbrack h_y(t)\,cos\phi(t) -h_x(t)\,sin\phi(t) \right\rbrack.
\label{eq:thermal_parts_phi}
\end{eqnarray}
\noindent
To derive the thermal torque, we have used the following identities.
\begin{subequations}
\begin{eqnarray}
\mathbf{\hat{e}_x} &=& sin\theta(t)\,cos\phi(t)\,\mathbf{\hat{e}_r} + cos\theta(t)\,cos\phi(t)\,\mathbf{\hat{e}_\theta} - sin\phi(t)\,\mathbf{\hat{e}_\phi},\\
\mathbf{\hat{e}_y} &=& sin\theta(t)\,sin\phi(t)\,\mathbf{\hat{e}_r} + cos\theta(t)\,sin\phi(t)\,\mathbf{\hat{e}_\theta} + cos\phi(t)\,\mathbf{\hat{e}_\phi},\\
\mathbf{\hat{e}_z} &=& cos\theta(t)\,\mathbf{\hat{e}_r} - sin\theta(t)\,\mathbf{\hat{e}_\theta}.
\end{eqnarray}
\end{subequations}
\begin{subequations}
\begin{eqnarray}
\mathbf{\hat{e}_r} \times \mathbf{\hat{e}_x} &=& cos\theta(t)\,cos\phi(t)\,\mathbf{\hat{e}_\phi} + sin\phi(t)\,\mathbf{\hat{e}_\theta}, \\
\mathbf{\hat{e}_r} \times \mathbf{\hat{e}_y} &=& cos\theta(t)\,sin\phi(t)\,\mathbf{\hat{e}_\phi} - cos\phi(t)\,\mathbf{\hat{e}_\theta}, \\
\mathbf{\hat{e}_r} \times \mathbf{\hat{e}_z} &=& - sin\theta(t)\,\mathbf{\hat{e}_\phi}.
\end{eqnarray}
\end{subequations}

The magnetization dynamics under the action of these two torques $\mathbf{T_{E}}(t)$ and 
$\mathbf{T_{TH}}(t)$ is described by the stochastic Landau-Lifshitz-Gilbert (LLG) equation as follows.
\begin{equation}
\cfrac{d\mathbf{n_m}(t)}{dt} - \alpha \left(\mathbf{n_m}(t) \times \cfrac{d\mathbf{n_m}(t)}
{dt} \right) = -\cfrac{|\gamma|}{M_V} \left\lbrack \mathbf{T_E}(t) +  \mathbf{T_{TH}}(t)\right\rbrack.
\label{LLG}
\end{equation}

In the spherical coordinate system with constant magnitude of magnetization, the following relation holds:
\begin{equation}
\cfrac{d\mathbf{n_m}(t)}{dt} = \cfrac{d\theta(t)}{dt} \, \mathbf{\hat{e}_\theta} + sin \theta (t)\, \cfrac{d\phi(t)}{dt}
\,\mathbf{\hat{e}_\phi}.
\end{equation}
Accordingly,
\begin{equation}
\alpha \left(\mathbf{n_m}(t) \times \cfrac{d\mathbf{n_m}(t)}{dt} \right) = - \alpha sin 
\theta(t) \, \phi'(t) \,\mathbf{\hat{e}_\theta} +  \alpha \theta '(t) \, \mathbf{\hat{e}_\phi}
\end{equation}
where ()' denotes $d()/dt$. This allows us to write
\begin{multline}
\cfrac{d\mathbf{n_m}(t)}{dt} - \alpha \left(\mathbf{n_m}(t) \times \cfrac{d\mathbf{n_m}(t)}
{dt} \right) = (\theta '(t) + \alpha sin \theta (t)\, \phi'(t)) \, \mathbf{\hat{e}_\theta} 
+ 
(sin \theta(t) \, \phi ' (t) - \alpha \theta '(t)) \,\mathbf{\hat{e}_\phi}.
\end{multline}
Equating the $\hat{e}_\theta$ and $\hat{e}_\phi$ components in both sides of Equation
(\ref{LLG}), we get
\begin{equation}
\theta ' (t) + \alpha sin \theta(t) \, \phi'(t)  
=  \frac{|\gamma|}{M_V} \, \left( B_{shape,\phi}(t) sin\theta(t) + P_\phi(t)\right),
\label{eq:LLG_separate_theta}
\end{equation}
\begin{equation}
sin \theta (t) \, \phi '(t) - \alpha \theta '(t) = \frac{|\gamma|}{M_V} 
\left( 2 B (t)sin\theta(t) cos\theta(t) - P_\theta(t)\right).
\label{eq:LLG_separate_phi} 
\end{equation}
Solving the above equations, we get the following coupled equations for the dynamics of $\theta(t)$ and $\phi(t)$.
\begin{eqnarray}
\left(1+\alpha^2 \right) \cfrac{d\theta(t)}{dt} &=& \frac{|\gamma|}{M_V} \left\lbrack 
 B_{shape,\phi}(t) sin\theta(t) - 2\alpha B(t) sin\theta (t)cos\theta (t) + \left(\alpha P_\theta + P_\phi \right) \right\rbrack.
 \label{eq:theta_dynamics}\\
\left(1+\alpha^2 \right) \cfrac{d \phi(t)}{dt} &=& \frac{|\gamma|}{M_V} 
\left\lbrack \alpha B_{shape,\phi}(t) + 2 B(t) cos\theta(t) - \frac{1}{sin\theta(t)} \left(P_\theta - \alpha P_\phi \right) \right\rbrack. \quad
	(sin\theta \neq 0)
\label{eq:phi_dynamics}
\end{eqnarray}

\section{\label{sec:initial_thermal} Initial fluctuation due to thermal torque}

We can see from Equation (\ref{shape-anisotropy}) that because $N_{d-zz} < N_{d-yy} \ll N_{d-xx}$, the shape 
anisotropy energy is minimum when $\theta = 0^{\circ}$ or 180$^{\circ}$.
Therefore, the magnetization of the unstressed magnet should be lined up along the easy axis ($z$-axis) in the 
absence of thermal perturbation. If that happens, then no amount of stress can budge the magnetization vector from
this orientation since the torque due to stress vanishes when $sin \theta = 0$ (see Equation (\ref{eq:stress_torque})).
Accordingly, $\theta = 0^{\circ}$ and 180$^{\circ}$ are called {\it stagnation points} since the 
magnetization vector stagnates 
at these locations.
In this section, we show mathematically that thermal torque can overcome stagnation 
and deflect the magnetization vector from the easy axis ($sin \theta = 0$). 

When $sin\theta=0$ and no stress is applied on the nanomagnet, Equations~(\ref{eq:LLG_separate_phi}) and~(\ref{eq:LLG_separate_theta}) yield
\begin{eqnarray}
\theta ' (t) &=&  \cfrac{|\gamma|}{M_V}\,P_\phi(t) \label{eq:initial_p_phi}\\
\alpha \theta '(t) &=& \cfrac{|\gamma|}{M_V}\,P_\theta(t). \label{eq:initial_p_theta}
\end{eqnarray}
\noindent
Substituting for $P_\theta(t)$ and $P_\phi(t)$ from Equations~(\ref{eq:thermal_parts_theta}) and~(\ref{eq:thermal_parts_phi}), and using $\theta=180^\circ$, we get
\begin{equation}
\alpha h_x(t) sin\phi(t) - \alpha h_y(t) cos\phi(t) = h_x(t) cos\phi(t) + h_y(t) sin\phi(t)
\end{equation}
which gives the expression for $\phi(t)$:
\begin{equation}
\phi(t) = tan^{-1} \left( \frac{\alpha h_y(t) + h_x(t)}{\alpha h_x(t)-h_y(t)} \right).
\label{eq:phi_t_thermal}
\end{equation}
If we substitute this value of $\phi(t)$ in Equation~(\ref{eq:initial_p_phi}) or in 
Equation~(\ref{eq:initial_p_theta}), we get  
\begin{equation}
\theta'(t) = -|\gamma| \frac{h_x^2(t) + h_y^2(t)}{\sqrt{(\alpha h_x(t)-h_y(t))^2 + (\alpha h_y(t) + h_x(t))^2}}.
\end{equation}
\noindent
We can see clearly from the above equation that thermal torque can deflect the magnetization axis 
when it is \emph{exactly} along the easy axis since the time rate of change of $\theta(t)$ 
is non-zero. Note that the initial deflection from the easy axis due to the thermal torque does not depend on the 
component of the random thermal field along the $z$-axis $h_z(t)$, which is a consequence 
of having $\pm$$z$-axis as the easy axes of the nanomagnet. However, once the magnetization 
direction is even slightly deflected from the easy axis, all  three components of the 
random thermal field along the $x$-, $y$-, and $z$-direction would come into play and affect the deflection.

\section{\label{sec:material_parameters}Material parameters}

The material parameters that characterize the magnetostrictive layer are given in  Table~\ref{tab:material_parameters}~\cite{RefWorks:179,RefWorks:176,RefWorks:178, materials}. 

\begin{table}[htbp]
\centering
\begin{tabular}{cc}
\hline \hline
& Terfenol-D\\
\hline \hline
Major axis (a) & 100 nm \\
Minor axis (b) & 90 nm \\
Thickness (t) & 6 nm \\
Young's modulus (Y) & 8$\times$10$^{10}$ Pa  \\
Magnetostrictive coefficient ($(3/2)\lambda_s$) & +90$\times$10$^{-5}$ \\
Saturation magnetization ($M_s$) & 8$\times$10$^5$ A/m \\
Gilbert's damping constant ($\alpha$) & 0.1 \\
\hline\hline
\end{tabular}
\caption{\label{tab:material_parameters}Material parameters for Terfenol-D as magnetostrictive layer.}
\end{table}

For the piezoelectric layer, we use lead-zirconate-titanate (PZT). The PZT layer is assumed to be four times thicker than the magnetostrictive layer so that any strain generated in it is transferred almost completely to the magnetostrictive layer. We will assume that the maximum strain that can be generated in the PZT layer is 500 ppm \cite{RefWorks:170}, which would require a voltage of 111 mV because $d_{31}$=1.8e-10 m/V for PZT \cite{pzt2}. The corresponding stress is the product of the generated strain ($500\times10^{-6}$) and the Young's modulus of the  magnetostrictive layer. So the maximum stress that can be generated on the Terfenol-D layer is 40 MPa.

 \section{\label{sec:results}Simulation results}
\begin{figure}[h]
\centering
\subfigure[]{\label{fig:thermal_theta_distribution_terfenolD_1000ns}\includegraphics[width=3.2in]
{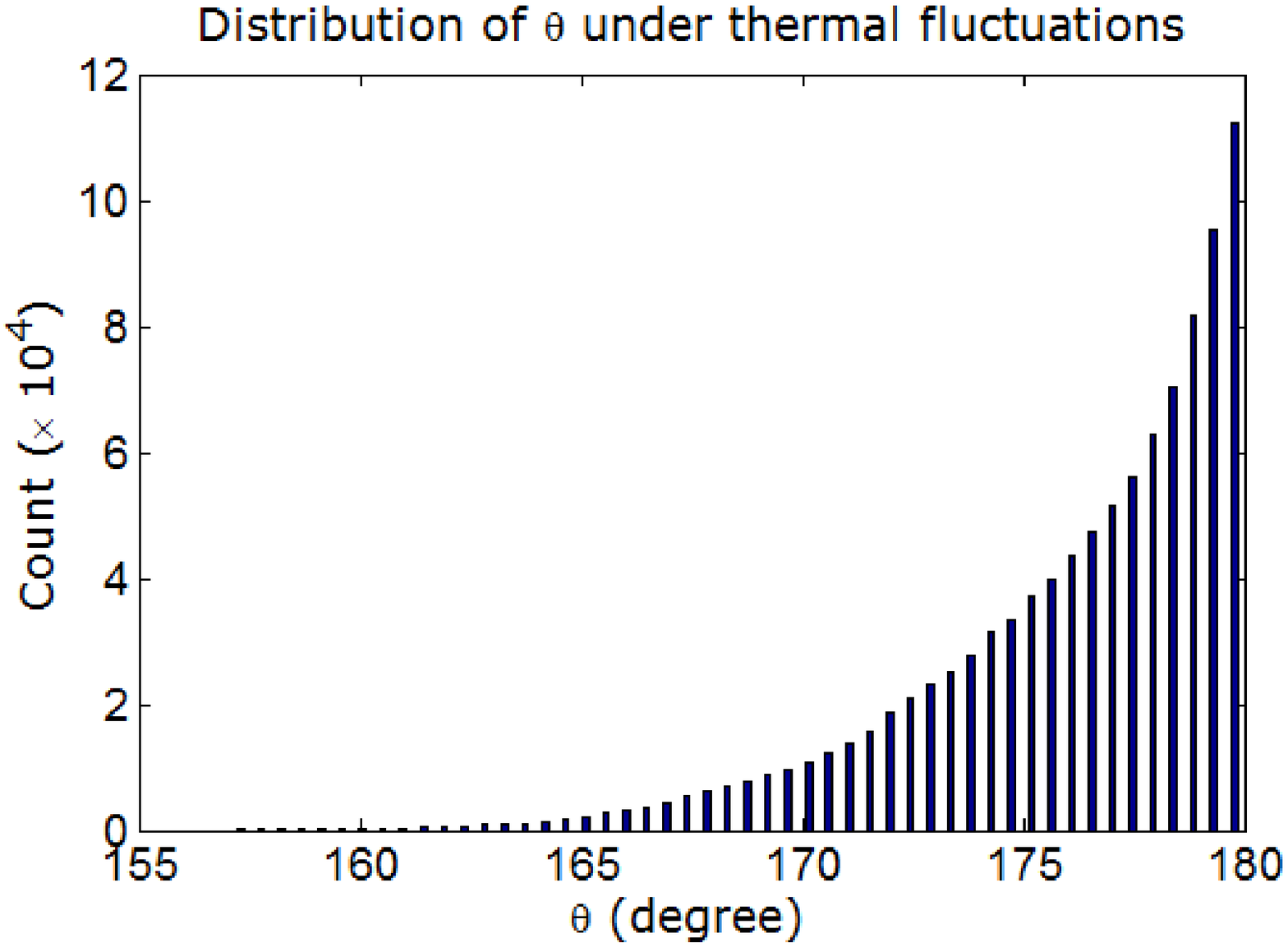}}
\subfigure[]{\label{fig:thermal_phi_distribution_terfenolD_1000ns}\includegraphics[width=3.2in]
{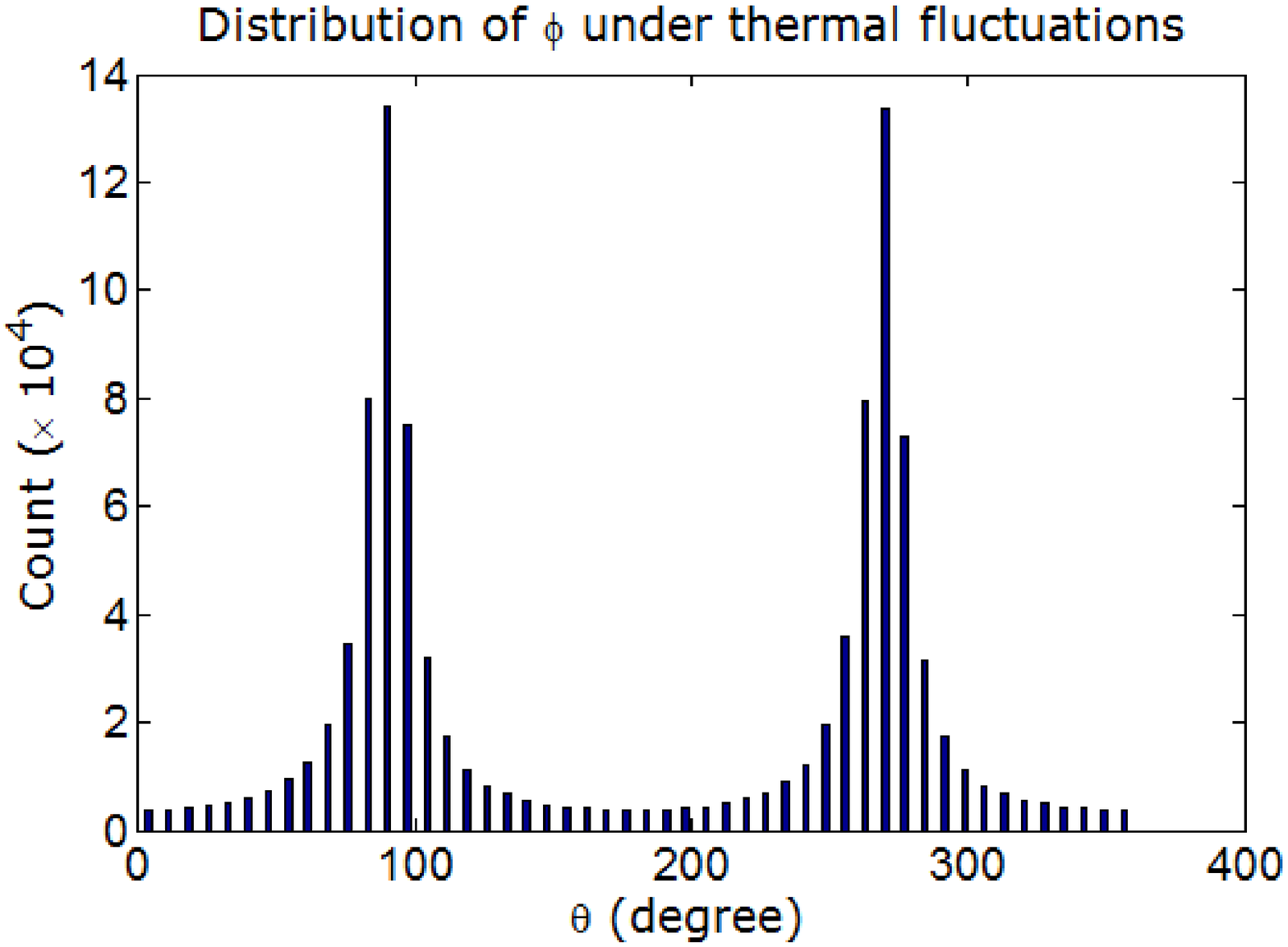}}
\caption{\label{fig:thermal_theta_phi_distribution_terfenolD_1000ns} Distribution of polar angle $\theta_{initial}$ and azimuthal angle $\phi_{initial}$ 
due to thermal fluctuations at room temperature (300 K). (a) Distribution of the polar angle $\theta_{initial}$. The mean of the distribution is $\sim$175$^\circ$, while the most likely value is 180$^{\circ}$.
 This is a nearly exponential distribution (Boltzmann-like). 
(b) Distribution of the azimuthal angle $\phi_{initial}$. These are two Gaussian distributions with peaks centered at $90^\circ$ and $270^\circ$ 
(or -90$^{\circ}$), which means that the most likely location of the 
 magnetization vector is in the plane of the nanomagnet.}
\end{figure}

\begin{figure}
\centering
\subfigure[]{\label{fig:dynamics_60ps_15MPa_theta_5deg_phi_90deg}\includegraphics[width=2.7in]
{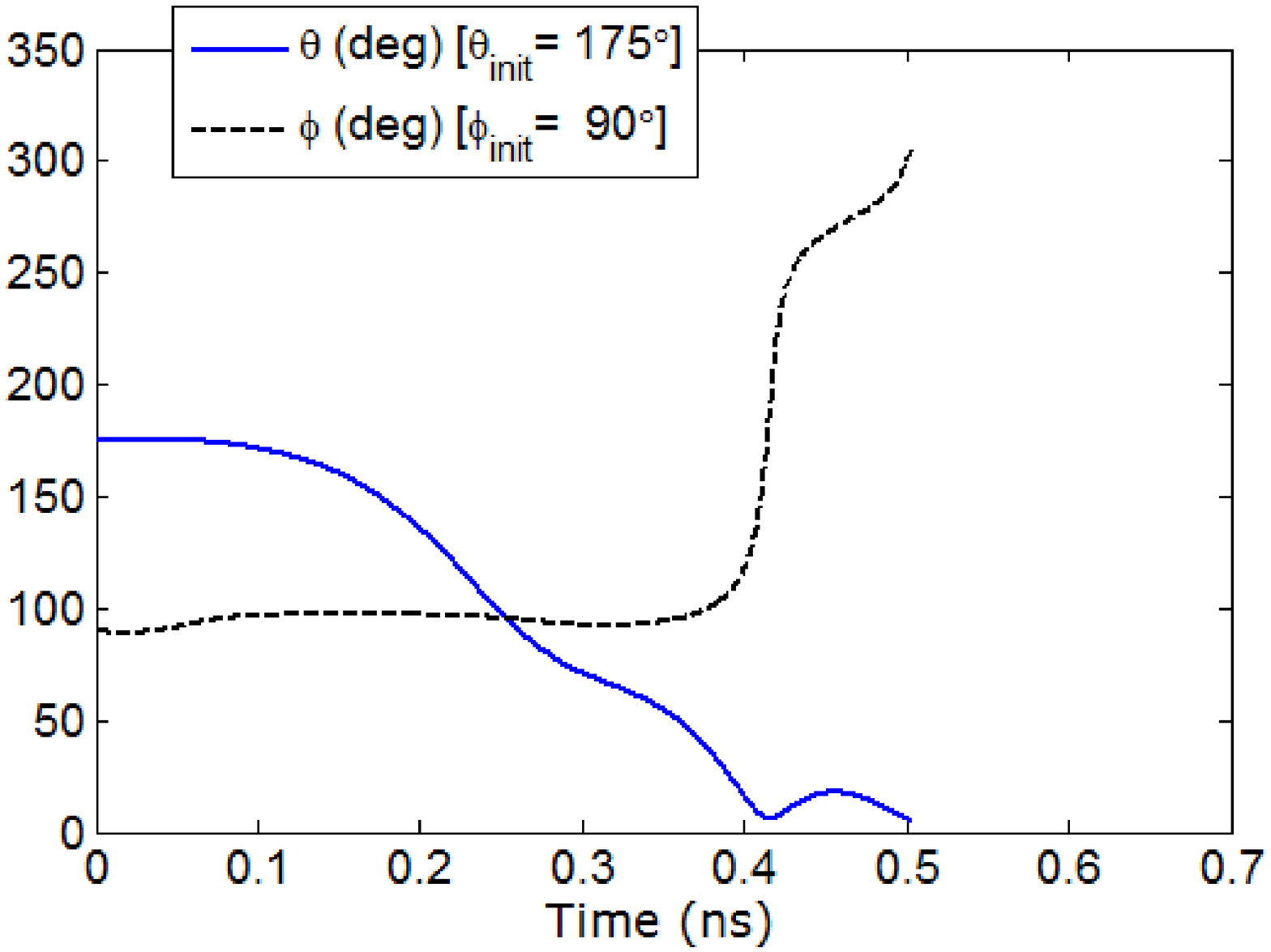}}
\subfigure[]{\label{fig:dynamics_60ps_15MPa_theta_5deg_phi_270deg}\includegraphics[width=2.7in]
{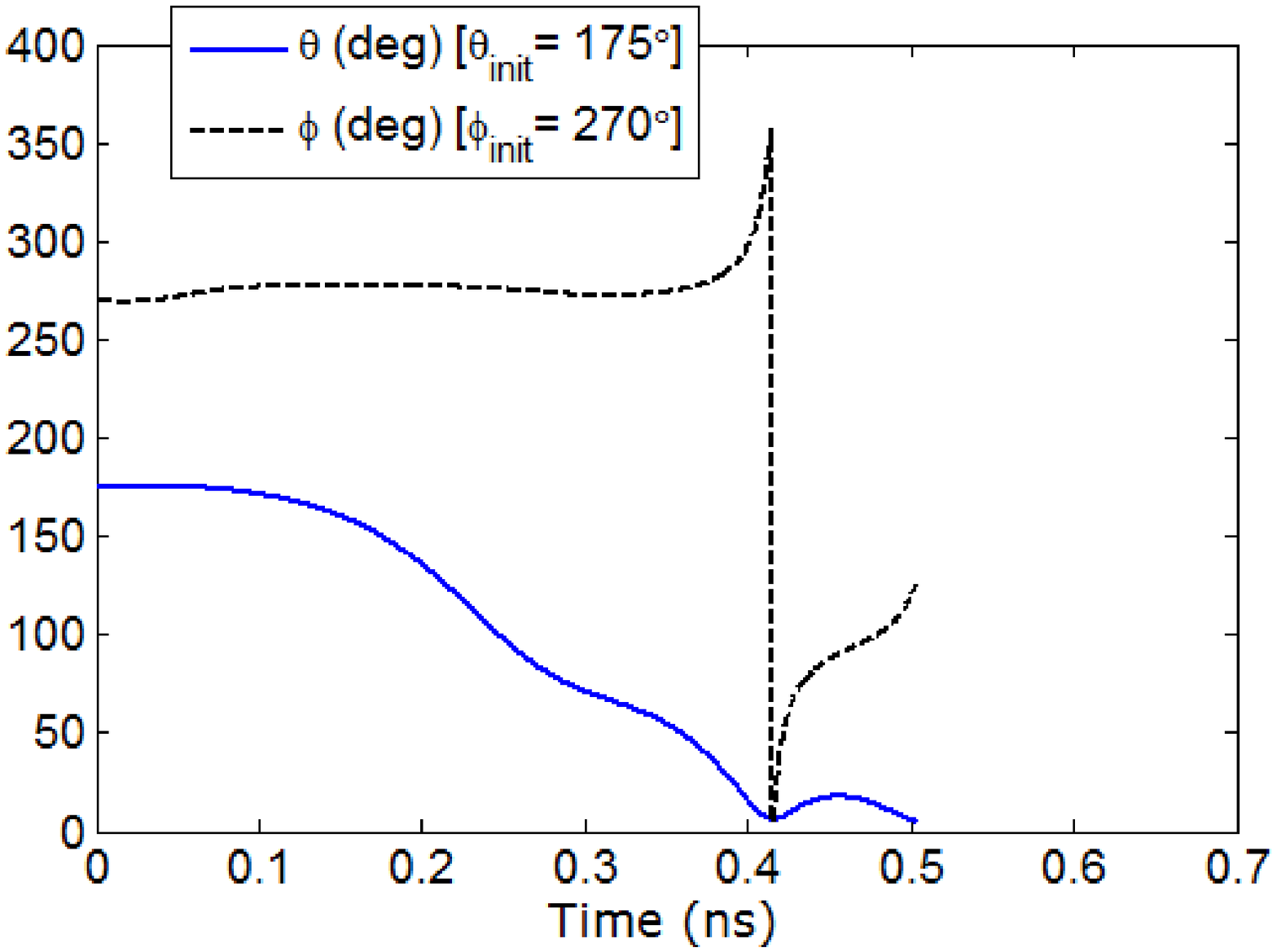}}
\subfigure[]{\label{fig:dynamics_60ps_15MPa_theta_5deg_phi_0deg}\includegraphics[width=2.7in]
{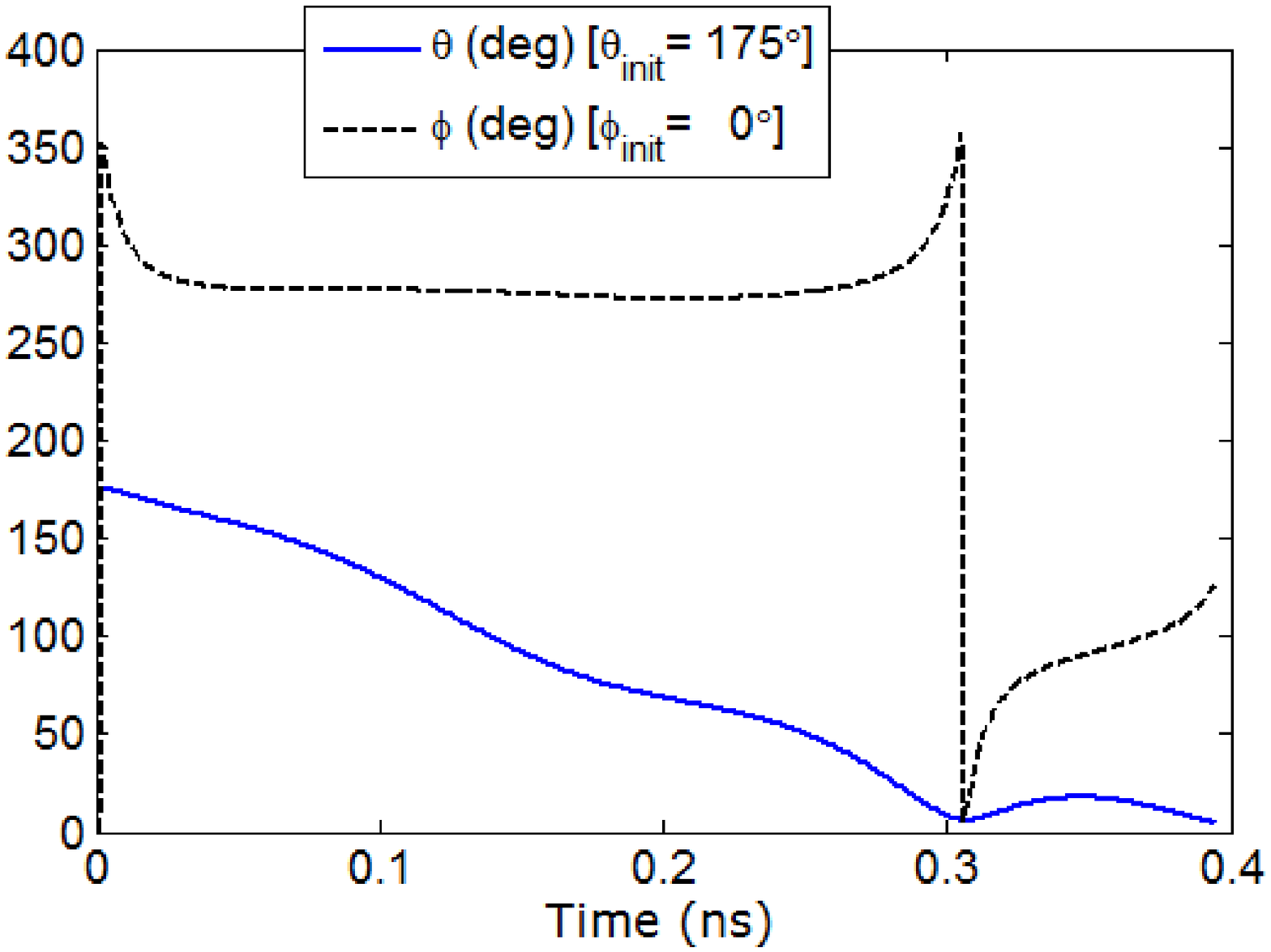}}
\subfigure[]{\label{fig:dynamics_60ps_15MPa_theta_5deg_phi_180deg}\includegraphics[width=2.7in]
{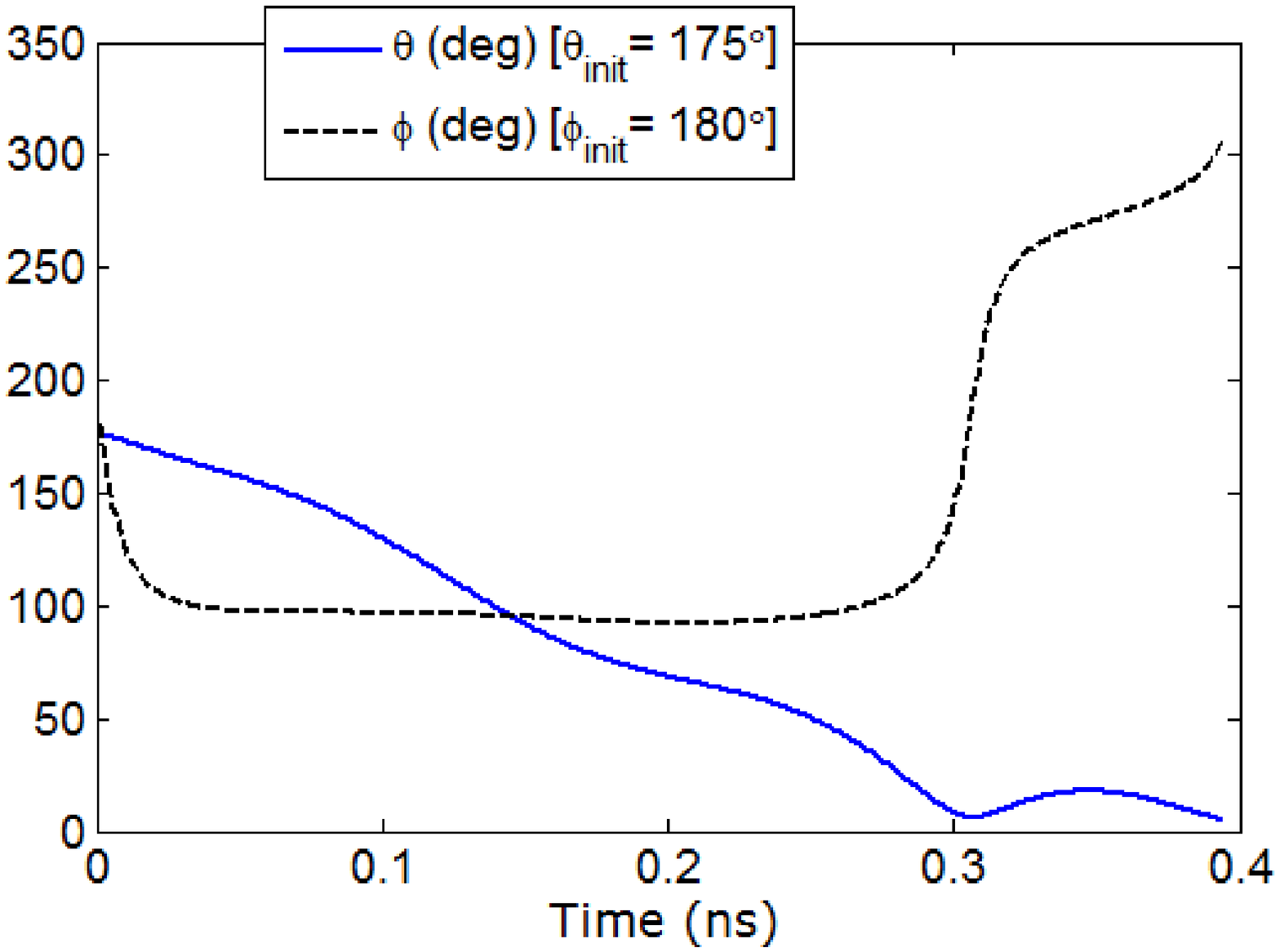}}
\caption{\label{fig:dynamics_60ps_15MPa_theta_5deg_xy} Temporal evolution of polar 
angle $\theta(t)$ and azimuthal angle $\phi(t)$ for a fixed $\theta_{initial}=175^\circ$ and four different values of 
$\phi_{initial}=\lbrace 90^\circ, 270^\circ, 0^\circ, 180^\circ \rbrace$. The applied stress is 15 MPa and the ramp duration is 60 ps. Thermal fluctuations have been ignored. (a) Temporal
evolution when $\theta_{initial}=175^\circ$, $\phi_{initial}=90^\circ$. (b) Temporal evolution when $\theta_{initial}=175^\circ$, $\phi_{initial}=270^\circ$. 
(c) Temporal evolution when $\theta_{initial}=175^\circ$, $\phi_{initial}=0^\circ$. (d) 
Temporal evolution when $\theta_{initial}=175^\circ$, $\phi_{initial}=180^\circ$. Note that when $\theta$ reaches $90^\circ$ or even earlier, 
$\phi$ always resides in the good quadrant [($90^\circ, 180^\circ$) or ($270^\circ, 360^\circ$)],
which makes the switching successful. 
For plots (a) and (b), the magnetization initially lies in the plane of the magnet ($\phi_{initial}=\pm90^\circ$) and precessional motion of magnetization due to applied stress is in the +$\hat{\mathbf{e}}_\phi$ direction, which increases $\phi$ with time. Thus, the magnetization starts out in the good quadrant. So both the motions (damped motion due to applied stress and the motion due to out-of-plane excursion depicted as $-2\alpha B(\phi)sin\theta cos\theta\,\hat{\mathbf{e}}_\theta$ and $-|B_{shape,\phi}(\phi)|sin\theta\,\hat{\mathbf{e}}_\theta$, respectively in the Fig.~3 of the main Letter) of magnetization are in the $-\hat{\mathbf{e}}_\theta$ direction so that $\theta$ decreases with time and the magnetization rotates in the correct direction towards $\theta = 90^{\circ}$.
The increasing out-of-plane excursion of the magnetization vector due to $\phi$ increasing with time
however is opposed by the damped motion due to out-of-plane excursion (depicted as $-\alpha |B_{shape,\phi}(\phi)|\,\hat{\mathbf{e}}_\phi$ in Fig.~3 of the main Letter), which tries to bring the magnetization back into the magnet's plane. 
These two effects quickly balance and $\phi$ assumes a stable value in the 
good quadrant as seen in the plots (the flat regions of the $\phi$-plots). When $\theta$ reaches $90^{\circ}$,
 the torque due to stress and shape anisotropy vanishes. At this point, we start to reverse the stress and the damped motion due to stress and shape anisotropy eventually becomes again in the $-\hat{\mathbf{e}}_\theta$ direction. That continues to rotate the magnetization in the right direction towards $\theta = 0^{\circ}$, ending in successful switching. Slightly past 0.4 ns, continuing $\phi$-rotation because of precessional motion due to stress and shape anisotropy pushes $\phi$ into the bad quadrant, but eventually it escapes into the other good quadrant. This brief excursion into the bad quadrant causes the ripple in $\theta(t)$ versus $t$.
For plots (c) and (d), the magnetization vector is initially lifted far out of the magnet's plane  ($\phi_{initial}=0^\circ,180^\circ$), where the huge out-of-plane shape anisotropy  cannot be overcome by the stress anisotropy and $B(\phi)$ becomes positive, i.e., $|B_{stress}| < |B_{shape}(\phi)|$. So magnetization precesses in the clockwise direction ($-\hat{\mathbf{e}}_\phi$) rather than in the anticlockwise direction ($+\hat{\mathbf{e}}_\phi$). Thus $\phi$ decreases with time, which immediately takes magnetization inside a good quadrant and eventually $|B_{stress}|$ becomes greater than $|B_{shape}(\phi)|$. Then $\phi$ assumes a stable value because of the damped motion due to out-of-plane excursion (depicted as $-\alpha |B_{shape,\phi}(\phi)|\,\hat{\mathbf{e}}_\phi$ in Fig.~3 of the main Letter). Afterwards, switching occurs similarly as for the cases (a) and (b). 
}
\end{figure}

\begin{figure}
\centering
\subfigure[]{\label{fig:thermal_distribution_delay90_60ps_15MPa}\includegraphics[width=3.2in]
{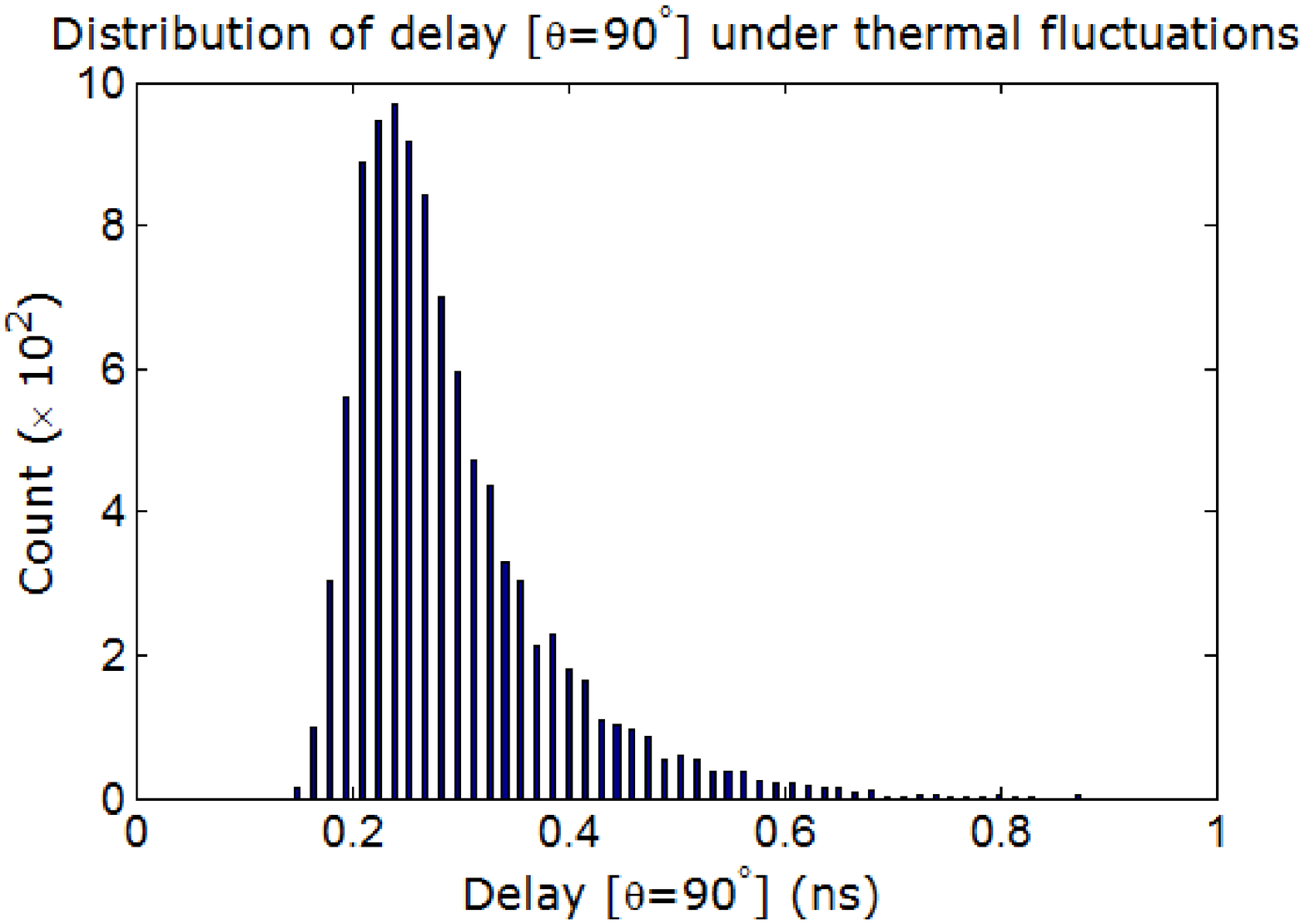}}
\subfigure[]{\label{fig:thermal_distribution_phi90_60ps_15MPa}\includegraphics[width=3.2in]
{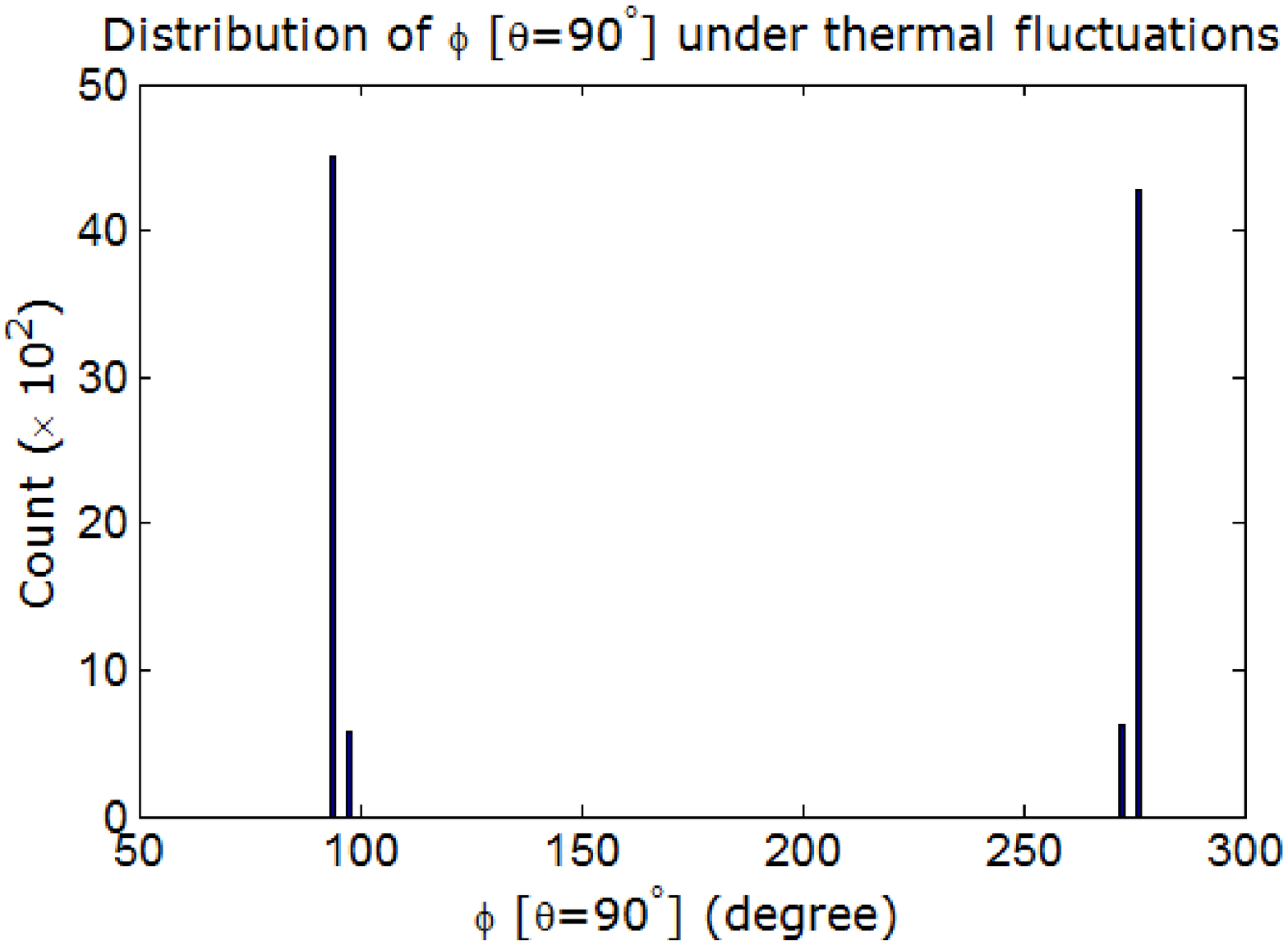}}
\caption{\label{fig:thermal_distribution_60ps_15MPa} Statistical distributions of
different quantities when 15 MPa stress is applied with 60 ps ramp duration on the magnet
at room temperature (300 K). (a) Distribution of time taken for $\theta$ to reach $90^\circ$ starting from ($\theta_{initial}$,$\phi_{initial}$)
where the latter are picked from the distributions in the presence of thermal fluctuations (see Fig.~\ref{fig:thermal_theta_phi_distribution_terfenolD_1000ns}). 
The wide distribution is caused by: (1) the initial angle distributions in Fig. \ref{fig:thermal_theta_phi_distribution_terfenolD_1000ns} 
and (2) thermal fluctuations during the transition from some $\theta = \theta_{initial}$  to $90^\circ$. 
We point out here that we do need a sensing circuitry to detect when $\theta$ reaches around 90$^{\circ}$, so that we can ramp down the stress thereafter. The sensing circuitry can be implemented with a spin-valve measurement 
of the magnetization. Such sensing circuitry is not uncommon in electronic circuits~\cite{yu06}. Some tolerance is nonetheless required since the sensing circuitry cannot be perfect. Our simulation shows that the internal dynamics works correctly as long as the stress is ramped down when $\theta$ is in the interval [85$^{\circ}$, 110$^{\circ}$], i.e. it does not have to be exactly 90$^{\circ}$. If magnetization reaches at $\theta=90^\circ$ (even past it) and stress is not withdrawn soon enough, then magnetization can end up at $\phi=\pm 90^\circ$ (potential energy minima), upon which the success probability would be 50\% since thermal fluctuations can put it in either direction of the potential landscape. Thus, we take advantage of such non-equilibrium scenario for which a sufficiently high magnitude of stress and a sufficiently fast ramp rate are necessary.
(b) Distribution of azimuthal 
angle $\phi$ when $\theta$ reaches $90^\circ$. Note that  $\phi$ always resides in the
good quadrant [($90^\circ, 180^\circ$) or ($270^\circ, 360^\circ$)] and has a fairly
narrow distribution. As explained in the main Letter, a  high stress and  fast ramp rate are required to ensure that $\phi$ is in the good quadrants, which is conducive to successful switching.}
\end{figure}

\begin{figure}
\centering
\subfigure[]{\label{fig:dynamics_failed_60ps_10MPa}\includegraphics[width=3.2in]
{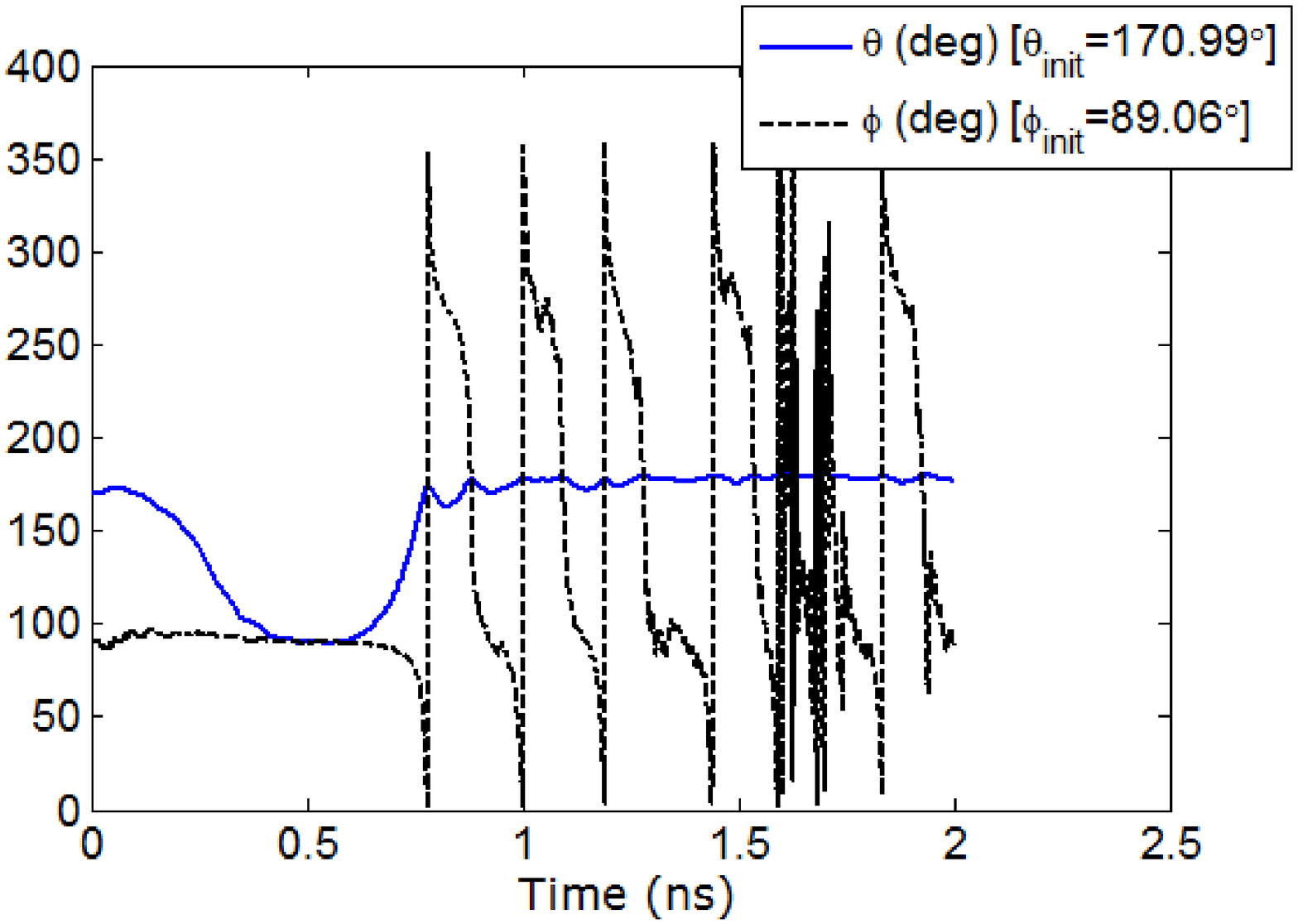}}
\subfigure[]{\label{fig:dynamics_failed_120ps_30MPa}\includegraphics[width=3.2in]
{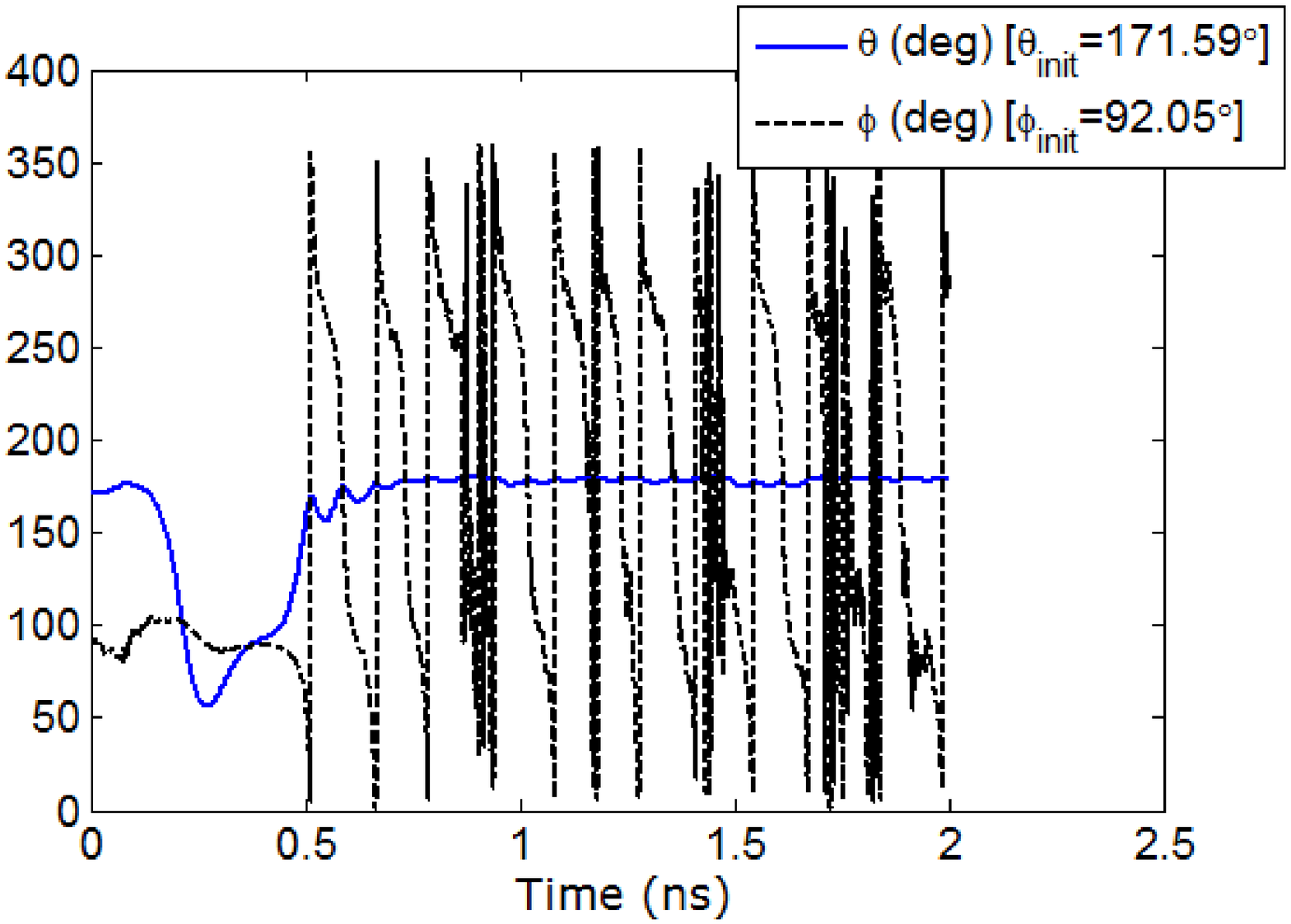}}
\caption{\label{fig:dynamics_failed} Temporal evolution of the polar 
angle $\theta(t)$ and azimuthal angle $\phi(t)$ when magnetization fails to switch and backtracks to the initial state. Simulations are 
carried out for room temperature (300 K). 
(a) The applied stress is 10 MPa and the ramp duration is 60 ps. When the polar angle $\theta$ reaches $90^\circ$, the 
azimuthal angle $\phi$ has ventured into the bad quadrant ($0^\circ$,90$^\circ$). Thus, switching eventually fails. 
(b) The applied stress is 30 MPa and the ramp duration is 120 ps. When the polar angle $\theta$ reaches $90^\circ$, the azimuthal angle 
$\phi$ is greater than $90^\circ$ and hence is in a good quadrant. However, after reaching $\theta\simeq50^\circ$, the magnetization 
backtracks to the initial state and switching fails. This happens because of the long ramp duration and during this passage of time thermal fluctuations have ample opportunity to scuttle the switching by bringing $\phi$ into the bad quadrant ($0^\circ$,90$^\circ$)}
\end{figure}

Some additional simulation results and corresponding discussions are given in 
Figures~\ref{fig:thermal_theta_phi_distribution_terfenolD_1000ns} -~\ref{fig:dynamics_failed}.

\clearpage
\pagebreak


\makeatletter 
\renewcommand\@biblabel[1]{[S#1]}
\makeatother

\clearpage
\pagebreak

%